\documentclass[acmtog,screen,authorversion]{acmart}
\usepackage{pdfpages}
\usepackage{cleveref}
\usepackage{nicefrac}
\usepackage{subcaption}
\usepackage{tabularx}
\usepackage{subfiles}

\usepackage{xcolor}

\crefformat{section}{Section~#2#1#3}
\crefformat{subsection}{Section~#2#1#3}
\crefformat{equation}{Eq.~#2#1#3}
\crefformat{figure}{Fig.~#2#1#3}
\crefformat{table}{Tab.~#2#1#3}

\AtBeginDocument{%
  }

\copyrightyear{2025} 
\acmYear{2025} 
\setcopyright{cc}
\setcctype{by}
\acmConference[SIGGRAPH Conference Papers '25]{Special Interest Group on
Computer Graphics and Interactive Techniques Conference Conference Papers
}{August 10--14, 2025}{Vancouver, BC, Canada}
\acmBooktitle{Special Interest Group on Computer Graphics and Interactive
Techniques Conference Conference Papers (SIGGRAPH Conference Papers '25),
August 10--14, 2025, Vancouver, BC, Canada}
\acmDOI{10.1145/3721238.3730593}
\acmISBN{979-8-4007-1540-2/2025/08}

\usepackage{colortbl}

\definecolor{yellow}{rgb}{1, 1, 0.7}
\definecolor{orange}{rgb}{1, 0.85, 0.7}
\definecolor{tablered}{rgb}{1, 0.7, 0.7}

\newcommand{\best}{\cellcolor{tablered}}
\newcommand{\sbest}{\cellcolor{orange}}
\newcommand{\tbest}{\cellcolor{yellow}}

\usepackage{expl3}
\ExplSyntaxOn
\NewDocumentCommand{\anonym}{ +m }
{
  \int_compare:nNnTF { \str_count:n {#1} } > { 0 }
  {
    \prg_replicate:nn { \str_count:n {#1} - 1 } { X\hspace{0pt plus 1pt} } X
  }
  {
    \mbox{} 
  }
}
\ExplSyntaxOff

\begin{document}

\title{Guiding-Based Importance Sampling for Walk on Stars}

\author{Tianyu Huang}
\affiliation{%
  \institution{School of Software and BNRist, Tsinghua University}
  \city{Beijing}
  \country{China}
}
\email{huang-ty21@mails.tsinghua.edu.cn}
\orcid{0009-0003-5405-4477}

\author{Jingwang Ling}
\affiliation{%
  \institution{School of Software and BNRist, Tsinghua University}
  \city{Beijing}
  \country{China}
}
\email{lingjw20@mails.tsinghua.edu.cn}
\orcid{0000-0001-8746-8578}

\author{Shuang Zhao}
\affiliation{%
 \institution{University of California Irvine}
 \city{Irvine}
 \country{United States of America}
}
\email{shz@ics.uci.edu}
\orcid{0000-0003-4759-0514}

\author{Feng Xu}
\authornote{Corresponding author.}
\affiliation{%
  \institution{School of Software and BNRist, Tsinghua University}
  \city{Beijing}
  \country{China}
}
\email{xufeng2003@gmail.com}
\orcid{0000-0002-0953-1057}

\renewcommand{\shortauthors}{Tianyu Huang, Jingwang Ling, Shuang Zhao, and Feng Xu}

\begin{abstract}
\emph{Walk on stars (WoSt)} has shown its power in being applied to Monte Carlo methods for solving partial differential equations, but the sampling techniques in WoSt are not satisfactory, leading to high variance.  
We propose a guiding-based importance sampling method to reduce the variance of WoSt. 
Drawing inspiration from path guiding in rendering, we approximate the directional distribution of the recursive term of WoSt using online-learned parametric mixture distributions, decoded by a lightweight neural field. This adaptive approach enables importance sampling the recursive term, which lacks shape information before computation.
We introduce a reflection technique to represent guiding distributions at Neumann boundaries and incorporate multiple importance sampling with learnable selection probabilities to further reduce variance. 
We also present a practical GPU implementation of our method. 
Experiments show that our method effectively reduces variance compared to the original WoSt, given the same time or the same sample budget. 
Code and data for this paper are at \url{https://github.com/tyanyuy3125/elaina}.
\end{abstract}

\ccsdesc[500]{Mathematics of computing~Partial differential equations}
\ccsdesc[500]{Computing methodologies~Rendering}
\ccsdesc[500]{Mathematics of computing~Probabilistic algorithms}

\keywords{Importance sampling, walk on stars, Monte Carlo methods}

\begin{teaserfigure}
    \includegraphics[width=\linewidth]{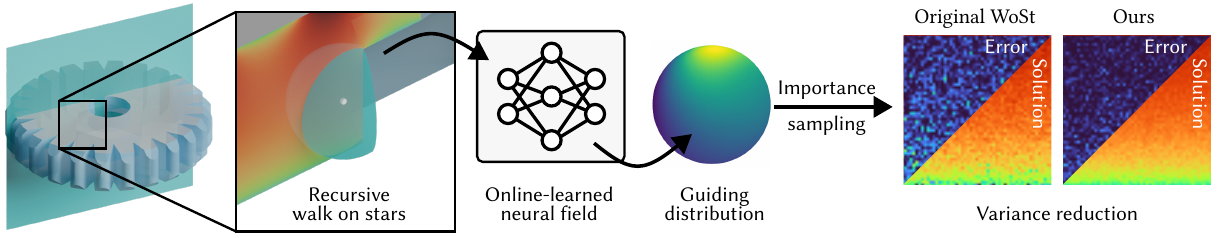}
    \centering
    \caption{An overview of our core method: We utilize an online-learned neural field to model the space-conditioned importance sampling distribution (guiding distribution) for the recursive term of the walk on stars (WoSt) estimator, achieving effective variance reduction.}
    \label{fig:rws_exp}
\end{teaserfigure}

\maketitle

\section{Introduction}
Partial differential equations (PDEs) are commonly used to model various phenomena such as heat transfer, wave propagation, and fluid dynamics in physics, engineering, and computer graphics.
Recently, Monte Carlo methods for solving PDEs have been gaining attention due to their advantages, such as eliminating the need for meshing, handling intricate geometries, and supporting local evaluations.
This series of methods began with \citet{sawhney2020mcgp}, who applied \emph{walk on spheres (WoS)}~\cite{muller1956wos} to the Dirichlet Poisson problem, which was later extended by \emph{walk on stars (WoSt)}~\cite{sawhney2023wost, miller2024robin} to support more boundary conditions.
These advancements have found broad applications, including volume rendering~\cite{qi22bidirectional}, fluid simulation~\cite{lavoie2022mcfluid, jain2024mcfluid}, heat simulation~\cite{delambilly2023heat}, robotics~\cite{muchacho2024pathplanning}, machine learning~\cite{nam2024solving}, shape modeling~\cite{degoes2024barycentric}, and infrared rendering~\cite{bati2023infrared}.

Unfortunately, WoSt suffers from high variance and slow convergence, especially for problems with complex domains or boundary conditions.
To address this problem, several recent methods~\cite{qi22bidirectional, li2023neuralcaches, bakbouk2023mvc, miller2023bvc} have adopted variance reduction techniques originated in Monte Carlo rendering~\cite{veach1995bdpt,muller2021nrc} for WoSt. %
The importance sampling family is effective for variance reduction in rendering, %
and so far, there have been explorations to importance sample the non-recursive Neumann and source contributions~\cite{sawhney2023wost, sawhney2020mcgp} of WoSt.
However, the recursive term of WoSt, lacking a closed form or shape information before computation, poses a challenge for traditional importance sampling techniques such as next event estimation~\cite{kajiya1986re, immel1986radiosity}. 
As it currently stands, the recursive term can only be sampled using uniform directional sampling.

We believe adaptive importance sampling techniques can handle the recursive term, as their online learning approach enables operation without prior knowledge of the distribution shape or modifications to the formulation.
Here, a representative approach in rendering is path guiding~\cite{mueller2017ppg, vorba2014gmm, herholtz2019vpg} which
uses the observations from earlier samples to optimize the importance sampling distribution (guiding distribution), and uses the distribution to generate subsequent samples.
Furthermore, recent advances using neural fields to encode space-conditioned parametric guiding distributions~\cite{dong2023npm, huang2024agpg} have demonstrated exceptional performance in variance reduction, indicating the potential to apply path guiding to WoSt for better reducing variance.

In this paper, we introduce a guiding-based importance sampling method to reduce the variance of WoSt. %
Specifically, we approximate the directional distribution of the recursive term of WoSt using von Mises-Fisher (vMF) mixtures. 
The mixtures are decoded by a lightweight neural field that takes space coordinates as input. 
Then we employ gradient-based online learning on the neural field to optimize the mixtures. 
This approach effectively handles high-frequency variations in the solution field.
Besides, we propose a sample reflection technique on the mixtures to represent guiding distributions at Neumann boundaries. 
This technique effectively reduces variance around Neumann boundaries compared to trivial approaches.
Additionally, we explore multiple importance sampling (MIS) with learnable selection probabilities in WoSt, which guarantees unbiasedness and adaptively adjusts the balance between guiding and uniform directional sampling, further reducing variance.
We present a practical GPU implementation 
of our method, enabling efficient parallelization including neural field training and geometric queries.

Our contributions are:
\begin{itemize}
    \item a guiding-based importance sampling method for the WoSt estimator,
    \item a sample reflection technique for guiding distributions at Neumann boundaries, and
    \item learnable selection probabilities to balance guiding and uniform directional sampling in WoSt.
\end{itemize}

These contributions address the unique challenges of WoSt, which are discussed in detail in \cref{sec:challenges}. We evaluate our method using several 2D and 3D problems under equal time or equal sample settings, demonstrating its effectiveness in variance reduction.

\section{Related Work}

\subsection{Monte Carlo Methods for Solving PDEs}
\paragraph{PDE Estimators and Their Applications} Recent exploration of Monte Carlo methods for solving PDEs~\cite{sawhney2024mcgpcourse} has gained significant attention in the graphics community, as they offer advantages over traditional methods like finite element (FEM) and finite difference (FDM) by circumventing spatial discretization challenges while providing better flexibility and performance.
The pioneering work, Monte Carlo Geometry Processing~\cite{sawhney2020mcgp} revisits \emph{walk on spheres (WoS)}~\cite{muller1956wos} to solve linear elliptic equations with Dirichlet boundary conditions, which is later extended to \emph{walk on stars (WoSt)}~\cite{sawhney2023wost, simonov2008wost, ermakov2009woh} to handle Neumann and Robin~\cite{miller2024robin} boundary conditions. Under the WoS(t) frameworks, the methods are further generalized to address problems with spatially varying coefficients~\cite{sawhney2022spatiallyvarying}, surface PDEs~\cite{sugimoto2024pwos}, and infinite domains~\cite{nabizadeh2021kelvin}. Monte Carlo methods have demonstrated broad applicability in both forward~\cite{jain2024mcfluid, lavoie2022mcfluid, delambilly2023heat, muchacho2024pathplanning, nam2024solving, degoes2024barycentric, bati2023infrared} and inverse~\cite{yu2024diffwos, miller2024differential, yilmazer2024inverse} PDE problems. In parallel with WoS(t), \emph{walk on boundary}~\cite{sugimoto2023wob}, another Monte Carlo estimator for PDEs, has been revisited and applied to fluid simulations~\cite{sugimoto2024mcfluid}. 

\paragraph{Variance Reduction Techniques} Similar to Monte Carlo rendering, Monte Carlo PDE solvers face the challenges of slow convergence and high variance. Various methods have been proposed to address these issues. 
 Reverse walk splatting~\cite{qi22bidirectional} and neural caches~\cite{li2023neuralcaches} are effective but biased. Neural control variates~\cite{li2024ncv} achieve good wall-time performance, but incur substantial computational overhead, making them impractical for real-time visualization. Meanwhile, mean value caching~\cite{bakbouk2023mvc} and boundary value caching~\cite{miller2023bvc} adopt novel formulations to reduce variance, but their recursive terms are still sampled uniformly, thus orthogonal to our method. So far, no existing method has performed importance sampling on the recursive term of WoSt. Our method explores using online-learned mixtures to perform importance sampling on the recursive term to reduce variance, practically and unbiasedly.

\subsection{Path Guiding in Rendering}
In Monte Carlo rendering, path guiding is a data-driven adaptive importance sampling scheme which learns the importance sampling distribution (guiding distribution) from previous samples to improve subsequent samples, thus reducing variance. 
Research in path guiding primarily focuses on how to represent, store, and optimize the guiding distribution across spatial, and recently, temporal domain~\cite{dong2024efficient}. Early attempts in this field include constructing spatially cached histograms~\cite{jensen1995pm}, cones~\cite{hey2002pg} or Gaussian mixtures~\cite{vorba2014gmm}. A well-known recent work is Practical Path Guiding~\cite{mueller2017ppg}, which utilizes SD-trees to implement path guiding practical for production environments. Subsequent works have considered volume rendering~\cite{herholtz2019vpg}, caustics~\cite{li2022pathcut, fan2023pgspecular}, path space~\cite{reibold2018completepathguiding}, variance-aware sampling~\cite{rath2020vapg}, spatio-directional mixture models~\cite{dorik2022sdmm}, and differentiable rendering~\cite{fan2024drpg}. 
In the deep learning era, path guiding based on neural networks have also been explored, such as employing convolutional neural networks to reconstruct radiance fields~\cite{huo2020conv, zhu2021conv}, or using invertible neural networks to model complex distributions~\cite{mueller2019neural}. Recently, methods utilizing neural fields to encode parameterized guiding distributions~\cite{huang2024agpg, dong2023npm} have emerged as the state-of-the-art in the field of path guiding. Our work is inspired by path guiding in rendering, with unique improvements tailored for 2D and 3D PDE problems, including reflection transformations on the guiding distribution to handle Neumann boundaries and learnable selection probabilities to combine uniform directional sampling and guiding adaptively.

\subsection{WoSt-Specific Challenges Beyond Light Transport}
\label{sec:challenges}

\subsubsection{Parallax Property} In rendering, the sampling region changes smoothly as viewpoint shifts. In contrast, WoSt's star-shaped sampling region exhibits abrupt deformations around non-convex geometry, expanding or shrinking dramatically with small evaluation-point movements. This instability challenges traditional guiding structures~\cite{ruppert2020pavmm}. This explains our empirical choice of neural fields.

\subsubsection{Boundary Handling}
\label{sec:boundary_handling}
Rendering methods often exploit cosine-weighted BRDFs to bound sampling domains~\cite{diolatzis2020practical}. However, WoSt lacks a BRDF analogue, and its Neumann boundaries may produce sharp, high-value distributions at grazing angles---unlike rendering, where BRDFs typically diminish at such angles. This discrepancy explains our adoption of sample reflection at Neumann boundaries.

\subsubsection{Selection Probability} While learnable selection probability is common in rendering~\cite{mueller2019neural, diolatzis2020practical, huang2024agpg}, it is used to combine guiding distributions and BSDFs, and the suitable distribution to combine for WoSt has not been proposed. We observe that, as the solution field of linear elliptic equations is smooth, the distributions on many small star-shaped regions near Dirichlet boundaries or Neumann silhouettes are close to uniform distribution. This consideration motivates performing learnable selection with uniform sampling for WoSt.

\section{Background}

\subsection{Linear Elliptic Equations}

Walk on stars (WoSt) primarily targets linear elliptic equations, which encompass a wide variety of forms. However, the structure of WoSt remains largely consistent across different formulations. We thus focus on the most typical case---the Poisson equation with Dirichlet and Neumann boundary conditions:
\begin{equation}
\begin{aligned}
\Delta u(x) &= f(x) \quad \text{on } \Omega, \\
u(x) &= g(x) \quad \text{on } \partial \Omega_\text{D}, \\
\frac{\partial u(x)}{\partial n_x} &= h(x) \quad \text{on } \partial \Omega_\text{N},
\end{aligned}
\label{eq:linear_elliptic}
\end{equation}
where the boundary of the domain $\Omega \subset \mathbb{R}^d~(d=2,3)$ is partitioned into a Dirichlet part $\partial \Omega_\text{D}$ and a Neumann part $\partial \Omega_\text{N}$ with prescribed values $g$ and $h$ \textit{(resp.)}. $\Delta$ is the negative-semidefinite Laplacian, $u: \Omega \rightarrow \mathbb{R}$ is the unknown solution, and $f: \Omega \rightarrow \mathbb{R}$ is a source term.
This equation can describe, \textit{e.g.}, the steady-state temperature distribution, where $f$ represents the heat source or sink, $g$ corresponds to the temperature on the boundary, and $h$ denotes the heat flux on the boundary.

\subsection{The Walk on Stars (WoSt) Estimator}

\label{sec:the_wost_estimator}

\begin{figure}[htbp]
    \includegraphics[width=0.95\columnwidth]{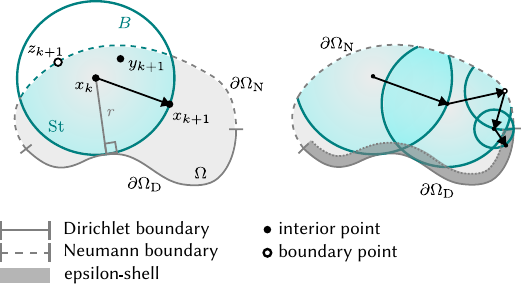}
    \centering
    \caption{Illustration of the WoSt estimator in 2D. \textit{Left}: An illustration of a \emph{step}. $x_k$ is the query point, $x_{k+1}$ is the next walk location, $y_{k+1}$ is the source sample point, and $z_{k+1}$ is the Neumann boundary sample point. \textit{Right}: An entire \emph{walk} of WoSt, which iteratively samples the next walk location until it reaches the Dirichlet boundaries' $\epsilon$-shell.}
    \label{fig:wost}
\end{figure}

We define the $\epsilon$-shell $\partial \Omega_\text{D}^\epsilon := \lbrace x \in \Omega : \min_{y \in \partial \Omega_\text{D}} \lVert x - y \rVert \leq \epsilon \rbrace$ for a manually specified small $\epsilon > 0$, which is the region where the walks terminate. The solution $u(x_k)$ at any query point $x_k \in \Omega\backslash\partial\Omega_\text{D}^\epsilon$ for \cref{eq:linear_elliptic} can be obtained using the following single-sample WoSt estimator~\cite{sawhney2023wost}:
\begin{equation}
\begin{aligned}
\langle u(x_k)\rangle &= \frac{P^B(x_k, x_{k+1}) \, \langle u(x_{k+1})\rangle}{\alpha(x_k) \, p^{\partial \text{St}(x_k, r)}(x_{k+1})} - \langle N\rangle + \langle S\rangle, \\
\langle N\rangle &= \frac{G^B(x_k, z_{k+1}) \, h(z_{k+1})}{\alpha(x_k) \, p^{\partial \text{StN}(x_k, r)}(z_{k+1})}, \\
\langle S\rangle &= \frac{G^B(x_k, y_{k+1}) \, f(y_{k+1})}{\alpha(x_k) \, p^{\text{St}(x_k, r)}(y_{k+1})},
\end{aligned}
\label{eq:wost_estm}
\end{equation}
where $\langle N\rangle$ and $\langle S\rangle$ are non-recursive Neumann and source contributions, \textit{resp.}; $B$ is a ball centered at $x_k$ with radius $r$; the radius $r$ equals to the smaller value between the distance from $x_k$ to $\partial\Omega_\text{D}$ and the distance from $x_k$ to the nearest silhouette on $\partial\Omega_\text{N}$; the star-shaped region $\mathrm{St} := B \cap \Omega$; $\partial \mathrm{St}$ denotes the boundaries of $\mathrm{St}$; $\partial \mathrm{StN}$ denotes the Neumann boundaries of $\mathrm{St}$; $\alpha(x_k)$ is set to $1$ if $x_k$ lies within $\mathrm{St}$, $\nicefrac{1}{2}$ if it lies on the boundary of $\mathrm{St}$, and $0$ if it lies outside $\mathrm{St}$; $G^B$ denotes the Green's function defined over the sphere $B$ ~\citep[Eq. 24]{sawhney2023wost}, while $P^B$, the Poisson kernel on $B$, is defined as $P^B = \nicefrac{\partial G^B}{\partial n}$; $p$ represents the PDF of the sampler. The illustration of this single-sample Monte Carlo estimator can be found in \cref{fig:wost}.

In a \emph{walk}, the recursive estimator begins from an arbitrary evaluation point within $\Omega \backslash \partial\Omega_\text{D}^\epsilon$; at each \emph{step}, the estimator performs up to three sampling operations around the query point $x_k$:

\begin{enumerate}
    \item Sample source contribution at point $y_{k+1} \in \mathrm{St}$.
    \item Sample Neumann contribution at point $z_{k+1} \in \partial\text{St} \cap \partial\Omega_\text{N}$.
    \item Sample the next walk location $x_{k+1} \in \partial\text{St}$.
\end{enumerate}

The walk continues until it reaches $\partial\Omega_\text{D}^\epsilon$, where it uses the Dirichlet data $g$ at the closest point $\bar{x}_k\in \partial \Omega_\text{D}$ to set $u(x_k) := g(\bar{x}_k)$. The first term of \cref{eq:wost_estm} is an unknown term that needs to be estimated recursively, similar to the scenario encountered in Monte Carlo rendering, thus we name it as \emph{recursive term}.
The recursive term lacks a closed form expression or shape information before computation, posing significant challenges for traditional importance sampling approaches.
Therefore, in this paper, we employ guiding-based importance sampling at this term to achieve variance reduction.

\subsection{von Mises-Fisher Mixture Model}
\label{sec:pmm}
To fit the target distribution (see \cref{sec:importance_sampling_recursive_term}) online for importance sampling, we adopt the von Mises-Fisher (vMF) mixture model. Since PDE problems are often solved in various dimensions, we adopt the generalized form of the vMF distribution. The vMF distribution on the $(d-1)$-sphere $\mathbb{S}^{d-1}$ in $\mathbb{R}^d~(d=2,3)$ is defined as:
\begin{equation}
v(\nu \mid \mu, \kappa) = \frac{\kappa^{d/2-1}}{(2\pi)^{d/2}I_{d/2-1}(\kappa)}\exp(\kappa \mu^\mathrm{T} \nu),
\end{equation}
where $\nu \in \mathbb{S}^{d-1}$ represents a direction, $\kappa \in [0,+\infty)$ and $\mu\in\mathbb{S}^{d-1}$ define the concentration and mean of the vMF distribution, \textit{resp.}, and $I_k$ denotes the modified Bessel function of the first kind at order $k$. We refer to Appendix A for specific forms of vMF distribution in different dimensions. The vMF mixture model is thus a convex combination of $K$ vMF components:
\begin{equation}
\mathcal{V}(\nu\mid\Theta)=\sum_{i=1}^{K} \lambda_i \cdot v (\nu \mid \mu_i, \kappa_i ),
\label{eq:vmm}
\end{equation}
where $\Theta\in\mathbb{R}^{(2+d)\times K}$ is the set of mixture parameters, containing $K$ vMF components, each with $\lambda_i \in [0,1]$, $\kappa_i$ and $\mu_i$. Here, the mixture weight $\lambda_i$ satisfies $\sum_i \lambda_i = 1$. The vMF mixture model is naturally defined on the sphere across various dimensions, which aligns with the directional nature of the recursive term.

For 3D problems, we use the sampling method for the mixtures from \citet{tokuyoshi2025vmf}, while for 2D problems, we use the method from \citet{best1979vmf}.

\section{Method}

\subsection{Importance Sampling the Recursive Term}

\label{sec:importance_sampling_recursive_term}

\begin{figure}[htbp]
    \includegraphics[width=0.85\columnwidth]{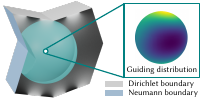}
    \centering
    \caption{Illustration of the guiding distribution at an arbitrary step. The guiding distribution is defined on $\mathbb{S}^{d-1}~(d=2,3)$, with a goal to be proportional to $\left|u\right|$ on $\partial \mathrm{St}$. The sampler uses this distribution for importance sampling.}
    \label{fig:target}
\end{figure}

We reparameterize the recursive term of the WoSt estimator (\cref{eq:wost_estm}) using the unit-length vector $\nu$, yielding an estimator of the following form:
\begin{equation}
\begin{aligned}
\langle u(x_k)\rangle &= \frac{\langle u(\nu; x_k)\rangle}{\lVert\mathbb{S}^{d-1}\rVert \, \alpha(x_k) \, p(\nu \mid x_k)} - \langle N\rangle + \langle S\rangle,
\end{aligned}
\end{equation}
where $\nu = \frac{x_{k+1} - x_k}{\lVert x_{k+1} - x_k \rVert} \in \mathbb{S}^{d-1}~(d = 2,3)$, $\lVert\mathbb{S}^{d-1}\rVert$ is the area of $\mathbb{S}^{d-1}$, and $p(\nu \mid x_k)$ is the directional PDF of the sampler \emph{conditioned on} $x_k$. 
We refer to Appendix B for a detailed derivation. 
For $x_k \in \partial\Omega_\text{N}$ with a normal\footnote{In this paper, we stipulate that the normal vector points toward the region where the walk is currently located. This assumption also applies to double-sided boundaries.} $n(x_k)$, the following condition of $p$ is satisfied, reflecting the fact that Neumann boundaries only have one valid side in random walk:
\begin{equation}
p(\nu \mid x_k) = 0 \quad \text{when } \nu \cdot n(x_k) \leq 0.
\label{eq:angular_req}
\end{equation}
In this formulation, the PDF $p_\text{u}$ of uniform directional sampling\footnote{The WoSt paper refers to this approach as \emph{importance sampling the Poisson kernel}. To avoid confusion with our method, we consistently refer to the sampling approach used in the original WoSt as \emph{uniform directional sampling} throughout this paper.} adopted by the original WoSt is thus:
\begin{equation}
p_\text{u}(\nu \mid x_k) =
\begin{cases} 
\frac{2 H(\nu \cdot n(x_k))}{\lVert \mathbb{S}^{d-1} \rVert}, & \text{if } x_k \in \partial\Omega_\text{N}, \\
\frac{1}{\lVert \mathbb{S}^{d-1} \rVert},  & \text{otherwise},
\end{cases}
\end{equation}
where $H(\cdot)$ is the unit step function. To reduce the variance of the recursive term, as illustrated in \cref{fig:target}, we aim to sample from an importance distribution defined on $\mathbb{S}^{d-1}$, referred to as the \emph{guiding distribution} $p_\text{g}$, that approximately satisfies

\begin{equation}
p_{\mathrm{g}}(\nu \mid x_k) \propto \left|u(\nu; x_k)\right|.
\label{eq:guiding_dist}
\end{equation}
We denote the right-hand side of \cref{eq:guiding_dist} as $p_\text{t}(\nu \mid x_k)$, \emph{i.e.}, the \emph{target distribution}. Next, we will discuss how to fit the guiding distribution to the target distribution.

\subsection{Representing and Learning the Guiding Distribution}
\label{sec:guiding_distribution}

\begin{figure}[htbp]
    \includegraphics[width=0.95\columnwidth]{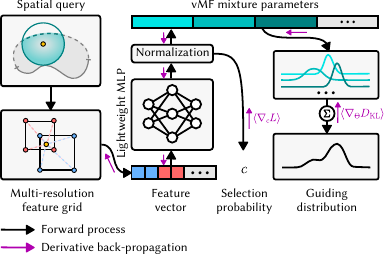}
    \centering
    \caption{Our neural field inference and training pipeline. During inference, the spatial query coordinates are encoded by a multi-resolution feature grid to produce feature vectors. These vectors are then fed into a lightweight MLP to obtain the parameters of the guiding distribution $\hat{\Theta}$ (\cref{eq:nn_to_theta}) and a selection probability $c$ (\cref{eq:mis_pdf}). During training, $\langle\nabla_\Theta D_\mathrm{KL}\rangle$ (\cref{eq:one_sample_kld}) for the guiding distribution and $\langle\nabla_c L\rangle$ (\cref{eq:one_sample_sp}) for the selection probability are back-propagated all the way to the grid and MLP parameters.}
    \label{fig:learning_overview}
\end{figure}

We adopt vMF mixture model (\cref{eq:vmm}) as the 
guiding distribution (\cref{eq:guiding_dist}) conditioned on $x \notin \partial\Omega_\text{N}$:
\begin{equation}
p_\text{g}(\nu\mid x) = \mathcal{V}(\nu\mid\Theta(x)).
\label{eq:guiding_dist_vmm}
\end{equation}
where $\Theta(x)$ is the space-conditioned form of $\Theta$ in \cref{eq:vmm}. To represent it, we employ a neural field $\mathbf{NN}(x\mid\Phi)$ with trainable parameters $\Phi$ to output its predicted values $\hat{\Theta}(x)$:
\begin{equation}
\mathbf{NN}(x\mid\Phi) = \hat{\Theta}(x),
\label{eq:nn_to_theta}
\end{equation}
where $\mathbf{NN}$ consists of a multi-resolution feature grid~\cite{hadadan2021nerad} and a lightweight multi-layer perceptron (MLP). The inference and training procedure is illustrated in \cref{fig:learning_overview}. Given a query point $x$, the neural field first encodes it through the multi-resolution feature grid, producing a feature vector that is then passed into the MLP. The MLP outputs a tensor of dimension $\left(\text{dim}\left(\Theta(x)\right) + 1\right)$, containing the unnormalized parameters $\hat{\Theta}'(x)$ of the mixtures, and the selection probability $c'$ (see \cref{sec:mis}). To ensure that each output component lies within a valid range, we apply a normalization mapping, as detailed in \cref{tab:reg_map}. The resulting mixtures with valid predicted parameters $\hat{\Theta}(x)$ are then used to sample the direction of the next walk location.

\begin{table}[htbp]
  \caption{Neural field outputs, and their corresponding normalization mappings. We refer to \cref{sec:pmm} for explanations of mixture parameters.}
  \label{tab:reg_map}
  \begin{tabular}{ccl}
        \toprule
        Parameter&$\mathbf{NN}$ output&Normalization mapping\\
        \midrule
        $\mu_i \in \mathbb{S}^{d-1}$&$\mu'_i \in \mathbb{R}^d$ & $\mu_i = {\mu'_i}/{\lVert\mu'_i\rVert}$\\
        $\kappa_i \in [0,+\infty)$&$\kappa'_i \in \mathbb{R}$ & $\kappa_i = \exp(\kappa'_i)$ \\
        $\lambda_i \in (0,1)$&$\lambda'_i \in \mathbb{R}$ & $\lambda_i = {\exp(\lambda'_i)}/{\sum_{j=1}^K \exp(\lambda'_j)}$\\ \midrule
        $c \in (0, 1)$ (\cref{eq:mis_pdf})&$c'\in \mathbb{R}$ & $c = \text{sigmoid}(c')$\\
      \bottomrule
    \end{tabular}
\end{table}

To fit the vMF mixture model $\mathcal{V}$ to the target distribution $p_\text{t}$ (see \cref{sec:importance_sampling_recursive_term}) at $x$, we introduce the Kullback-Leibler (KL) divergence as
\begin{equation}
D_{\mathrm{KL}}(p_\text{t}\parallel\mathcal{V};\Theta)=\int_\nu p_\text{t}(\nu)\,\mathrm{log}\frac{p_\text{t}(\nu)}{\mathcal{V}(\nu\mid\hat{\Theta})}\mathrm{d}\nu.
\label{eq:kld}
\end{equation}
The optimization objective of the neural-field parameters $\Phi$ is thus
\begin{equation}
\Phi^*=\mathop{\text{argmin}}_{\Phi}\mathbb{E}_x\Big[D_{\text{KL}}\Big(p_\text{t}(x)\parallel\mathcal{V};\Theta(x)\Big)\Big].
\label{eq:phi_estimate}
\end{equation}
We use a single-sample Monte Carlo estimator to estimate $\nabla_\Theta D_\text{KL}$:
\begin{equation}
\langle \nabla_\Theta D_\text{KL}(p_\text{t}\parallel\mathcal{V};\Theta)\rangle=-\frac{p_\text{t}(\nu) \nabla_\Theta \mathcal{V}(\nu \mid \hat{\Theta})}{\tilde{p}(\nu\mid\hat{\Theta})\mathcal{V}(\nu\mid\hat{\Theta})},
\label{eq:one_sample_kld}
\end{equation}
where $\tilde{p}$ is the PDF of the sampler given by \cref{eq:mis_pdf}. We back-propagate this derivative along the purple arrows in \cref{fig:learning_overview}, updating the parameters $\Phi$ of both the MLP and the multi-resolution feature grid using gradient-based optimization~\cite{kingma2014adam}.

\subsection{Sample Reflection at Neumann Boundaries}
For $x \in \partial\Omega_\mathrm{N}$, samples from the original mixtures span the full sphere thus do not satisfy \cref{eq:angular_req}. Therefore, we reflect invalid samples along the local tangent plane, resulting in the following PDF:
\begin{equation}
p_\text{g}(\nu \mid x) = 
\begin{cases} 
0, & \text{if } \nu \cdot n(x) \leq 0, \\
\mathcal{V}(\nu^{+}\mid\Theta(x)) + \mathcal{V}(\nu^{-}\mid\Theta(x)), & \text{otherwise},
\end{cases}
\end{equation}
where $\nu^{+} = \nu$, and $\nu^{-}$ is the reflection of $\nu^{+}$ off $\partial\Omega_N$ with normal $n(x)$, i.e. $\nu^{-} = \nu^{+} - 2 \left(\nu^{+} \cdot n(x)\right) n(x)$.

We have discussed how the strategy we adopt differs from similar strategies in rendering and fits the PDE problems in \cref{sec:boundary_handling}. \cref{sec:abl_refl} gives an ablation.

\subsection{Multiple Importance Sampling with Learnable Selection Probabilities}
\label{sec:mis}

Linear elliptic equations often contain relatively smooth regions. In these regions, uniform directional sampling is sometimes the optimal approach, whereas learned vMF mixtures might yield less accurate approximation. Moreover, solely relying on the learned distribution for importance sampling is an unstable strategy, potentially introducing variance or even bias~\cite{art2000safemis}. To address this, we introduce a learnable multiple importance sampling (MIS) method. Our approach is based on the single-sample balance heuristic~\cite{veach1995combine}, with the following MIS PDF:
\begin{equation}
\tilde{p}(\nu \mid x)=c(x) p_\text{g}(\nu \mid x) + (1 - c(x)) p_\text{u}(\nu \mid x),
\label{eq:mis_pdf}
\end{equation}
where $c(x)$ is the learnable selection probability, decoded from neural field output (\cref{tab:reg_map}). Following \citet{mueller2019neural}, our method learns $c(x)$ with the following loss function:
\begin{equation}
L = e D_\text{KL}(p_\text{t} \parallel \tilde{p}) + (1 - e) D_\text{KL}(p_\text{t} \parallel p_\text{g}),
\end{equation}
where $e$ is a fixed fraction that we set to $0.2$. The single-sample Monte Carlo estimator for $\nabla_c L$ for back-propagation is:
\begin{equation}
\langle \nabla_c L \rangle = -\frac{e p_\text{t}(\nu) (p_\text{g}(\nu \mid x) - p_\text{u}(\nu \mid x))}{\tilde{p}^2(\nu \mid x)}.
\label{eq:one_sample_sp}
\end{equation}
We find that performing our learnable MIS in combination with uniform directional sampling is sufficient to achieve excellent performance, as demonstrated in \cref{sec:abl_mis}. Currently, there are no alternative importance sampling distributions for the recursive term of WoSt. We anticipate that if such distributions are developed, they could be integrated with our method using this approach. We refer to \cref{sec:limitations} for further discussion.

\section{Implementation Details}

\subsection{Wavefront-style Monte Carlo PDE Solver}
Our guiding-based method (see \cref{sec:guiding_distribution}) consists of neural field inference and training, which benefits from batched input for parallelization. 
A common strategy in rendering for efficient batching is the wavefront-style architecture~\cite{laine2013megakernel}, where rays are generated and processed in batches. 
In light of this, we implement a wavefront-style Monte Carlo PDE solver on GPU.

As illustrated in \cref{fig:wavefront}, at each step of WoSt, our wavefront-style solver is divided into three stages:  
\begin{enumerate}
    \item Logic Stage: The distances to the nearest Dirichlet boundary and the Neumann silhouette are queried to compute the ball radius $r$ (see \cref{sec:the_wost_estimator}). Walks are then partitioned based on whether they fall within the $\epsilon$-shell.
    \item Evaluation Stage: This stage resembles the Material Stage in wavefront-style renderers. For walks inside the $\epsilon$-shell, contributions are evaluated on the nearest Dirichlet boundary. For walks outside the $\epsilon$-shell, contributions from source and Neumann boundaries are evaluated.
    \item Walk Stage: Analogous to the Ray Cast Stage in rendering, this stage samples the next walk location and updates throughput accordingly.
\end{enumerate}
As each stage consists of multiple GPU kernels, we use a Structure-of-Arrays (SoA) memory layout to transfer data between kernels and employ thread-safe queues to manage tasks for each kernel.

\subsection{Network Design and Implementation}
\label{sec:network_design}
\paragraph{Neural Field Architecture} 
To effectively capture the spatial variation of the space-conditioned guiding distribution, we use a hybrid architecture combining learnable spatial embeddings and a small MLP to encode the parametric mixtures.
We define $L$ embedding grids $G_l$, where $l=1,2,\dots,L$, to form a multi-resolution representation. 
Each grid spans the $d$-dimensional space of the problem with a spatial resolution of $D_l^d$.
A learnable embedding vector $\omega\in\mathbb{R}^F$ is associated at each lattice point of $G_l$.
To retrieve the spatial embedding for a point $x$, we perform bi-linear interpolation at neighboring lattice points for each resolution, and concatenate the resulting embedding vectors to form the full embedding $G(x)$:
\begin{equation}
    G(x\mid\Phi)=\mathop{\oplus}_{l=1}^L \text{bilinear} \left( x, V_l[x] \right), \quad G : \mathbb{R}^d \to \mathbb{R}^{L\times F} (d=2,3),
\end{equation}
where $V_l[x]$ is the embedding vectors at the corners of the cell enclosing $x$ within $G_l$, and $\mathop{\oplus}$ is the concatenation operation. 
$G(x)$ is subsequently mapped by an MLP with 3 layers, each containing 64 neurons.
Trainable parameters $\Phi$ in \cref{eq:nn_to_theta} consist of the learnable spatial embeddings and MLP weights.
We implement the neural field based on \emph{tiny-cuda-nn}~\cite{tiny-cuda-nn}, adopting \texttt{DenseGrid} as the encoding method of the point coordinates. 
We employ \texttt{ReLU} as the activation function of the MLP, and the output is normalized by mapping in \cref{tab:reg_map}, producing $\hat{\Theta}(x)$ in \cref{eq:nn_to_theta} and $c(x)$ in \cref{eq:mis_pdf}. We refer to Appendix D for detailed configurations of the neural field.

\paragraph{Online Training Scheme} 
Similar to Monte Carlo rendering, walks are evaluated in batches.
Each batch consists of one entire walk per evaluation point.
Once all walks in a batch are completed, a training stage updates the neural field using the information gathered. The updated neural field is then used to sample the next batch. 
Once the walks per point (wpp) reaches a certain threshold, typically 256 wpp, the guiding distribution converges. At this time, we terminate the training process and use the neural field exclusively for inference, further improving performance.

\begin{figure}[htbp]
    \includegraphics[width=\columnwidth]{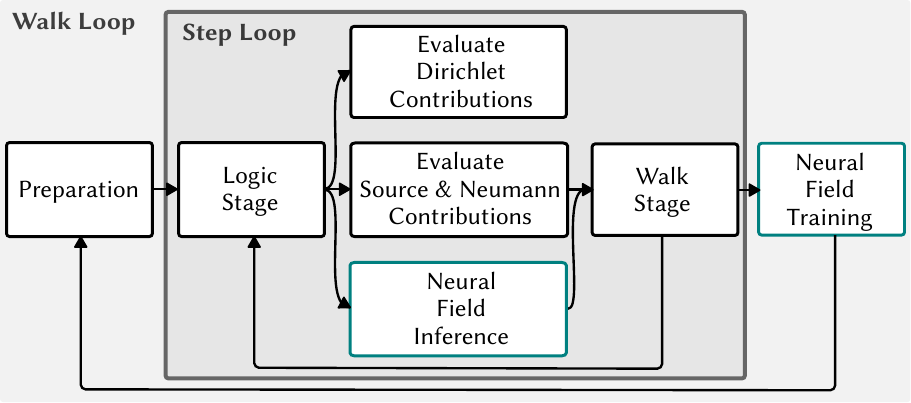}
    \centering
    \caption{Illustration of our wavefront-style Monte Carlo PDE solver including neural field inference and training stages.}
    \label{fig:wavefront}
\end{figure}

\paragraph{System Integration} As shown in \cref{fig:wavefront}, we insert the neural field inference stage before the walk stage and integrate the neural field training stage at the end of each walk loop. Our neural field and the WoSt integrator operate on separate CUDA streams, allowing the network inference and the evaluation stages to execute in parallel within the step loop. This approach maximizes the batch size during inference and training, avoiding the inefficiencies associated with single-sample inference or training.

\subsection{Geometric and Source Queries}
\label{sec:geometric_queries}
To demonstrate the practicality of our method, it is essential to implement a system with strong performance. The performance bottleneck of Monte Carlo PDE solvers typically lies in geometric queries. The authors of WoSt provide \texttt{Zombie}~\cite{Zombie} as the algorithm implementation and \texttt{FCPW}~\cite{FCPW} for geometric queries. However, \texttt{Zombie} is a CPU-only implementation, and although \texttt{FCPW} offers GPU support via Slang~\cite{he2018slang}, integrating Slang with \emph{tiny-cuda-nn} presents significant challenges. 

To address this, we develop a CUDA-based geometric and source query library. For geometric queries, we reference the implementation of WoBToolbox~\cite{sugimoto2023wob} and create a comprehensive query library for WoSt using a linear BVH~\cite{karras2012lbvh}. While \texttt{Zombie} uses a dense grid for source queries in 2D, which is memory inefficient if applied in 3D, we leverage NanoVDB~\cite{nanovdb}, a widely adopted GPU-based library for sparse volumetric storage and efficient source queries in 2D and 3D.

\section{Results and Discussion}

\label{sec:results}

Since we perform validation on visualization tasks and the experimental results exhibit large variations in scale, we employ the relative mean squared error (relMSE) as our quantitative metric. All experiments are performed on an AMD EPYC 9754 128-Core Processor with an RTX 4090D GPU. We set the hyperparameter $K=8$ except for \cref{sec:abl_vmf}. We refer to Appendix E for the complete quantitative results at 1024 wpp of all experiments.

Since our method is the first importance sampling approach targeting the recursive term of WoSt, we find it reasonable to validate its effectiveness by comparing with the original WoSt. In Appendix C, we provide additional comparisons with some existing variance reduction methods~\cite{qi22bidirectional, li2023neuralcaches}.

\subsection{Comparison with the Original WoSt}

\label{sec:comp_wost}

For a fair equal-time comparison, we implement both our method and the original WoSt on the same system, and disable other sampling techniques such as control variates, adaptive sampling, and stratified sampling. Russian roulette is enabled only when the walk length exceeds 128 to prevent infinite walks in scenarios involving Neumann boundaries. 

All experiments in this section are conducted with an evaluation grid of 1024$\times$1024, running 1024 walks per point (wpp). Our approach does not involve pretraining and uses all samples to form the final result, ensuring no additional samples are required compared to the baseline.

\subsubsection{3D Geometry}

\begin{figure*}
    \includegraphics[width=0.85\linewidth]{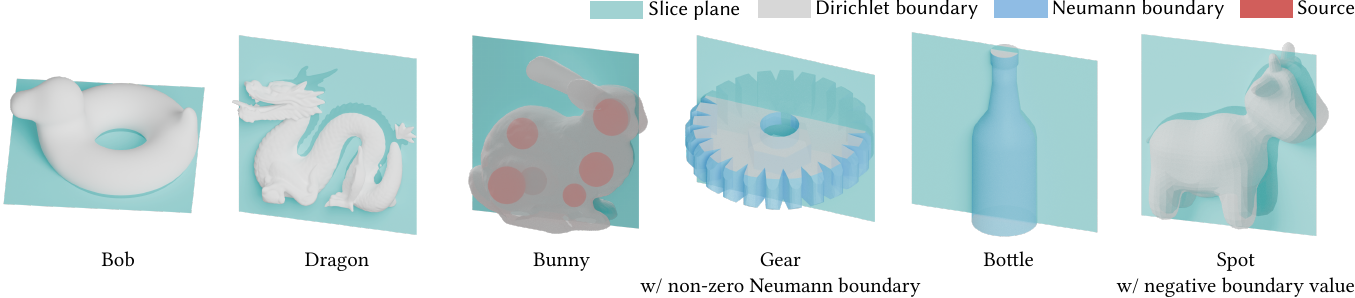}
    \centering
    \caption{Dataset for our 3D experiments. We adopt solving on slices~\citep[Section 5.2]{sawhney2020mcgp} as the visualization method.}
    \label{fig:3d_dataset}
\end{figure*}

We adopt solving on slices~\citep[Section 5.2]{sawhney2020mcgp} as the visualization method. For the dataset, following the experimental design of \citet{li2023neuralcaches}, we create six 3D problems to compare performance on Dirichlet boundaries (\cref{fig:3d_dataset}, columns 1–2), source (column 3), Neumann boundaries (columns 4–5), and problems involving signed values (column 6).
 
\begin{figure*}
    \includegraphics[width=0.95\linewidth]{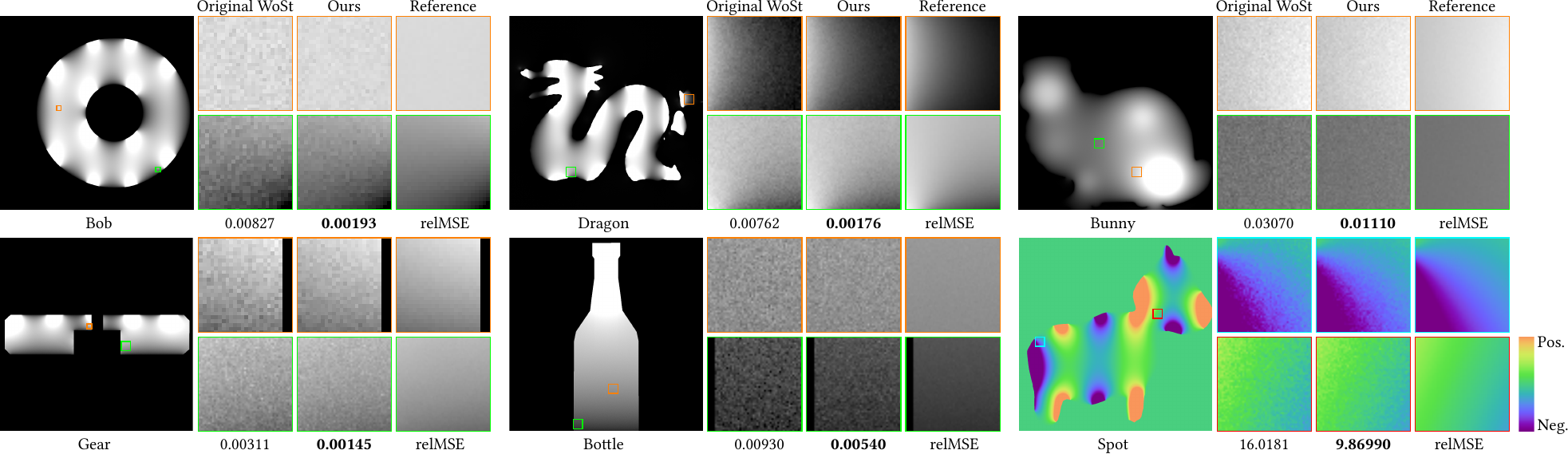}
    \centering
    \caption{Equal-sample qualitative results of our 3D experiments at 512 wpp. We follow \citet[Fig. 6]{sawhney2020mcgp}, using grayscale images to visualize the solution for positive-only results. Since WoSt typically exhibits variance as uniform salt-and-pepper noise~\citep[Section 4]{sawhney2023wost}, we zoom in on representative regions for visualization. We report relMSE at 512 wpp for each problem below the corresponding images.}
    \label{fig:exp_1}
\end{figure*}

\begin{figure*}
    \includegraphics[width=0.95\linewidth]{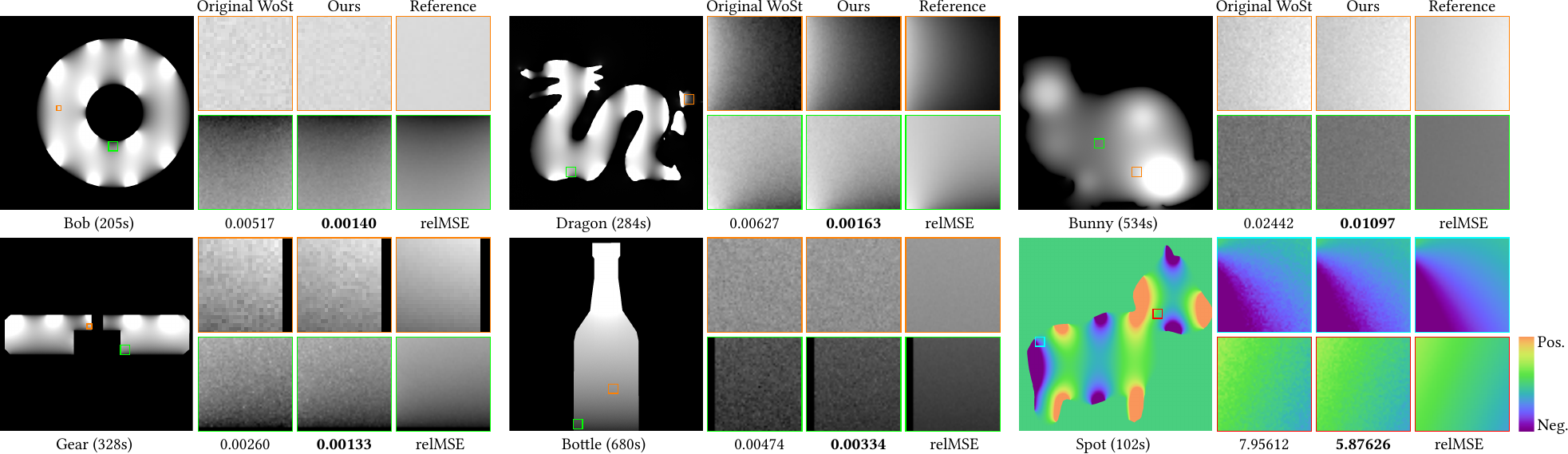}
    \centering
    \caption{Equal-time qualitative results of our 3D experiments. We report wall-time results for all 3D problems at their respective given runtime.}
    \label{fig:exp_1_equal_time}
\end{figure*}

\begin{figure*}
    \includegraphics[width=0.9\linewidth]{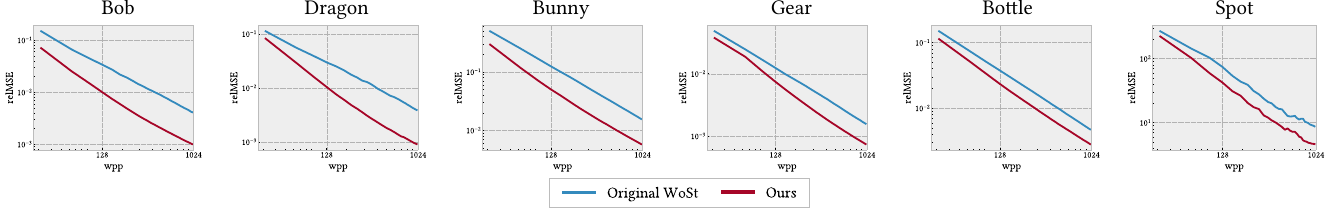}
    \centering
    \caption{relMSE plotted as a function of wpp of our 3D experiments. Under the condition of equal wpp, our method comprehensively surpasses the original WoSt in 3D problems. In the best scenario, it achieves more than a 4$\times$ reduction in variance, while in problems with relatively uniform solution distributions, it still delivers approximately a 2$\times$ improvement.}
    \label{fig:3d_wpp}
\end{figure*}

\begin{figure*}
    \includegraphics[width=0.9\linewidth]{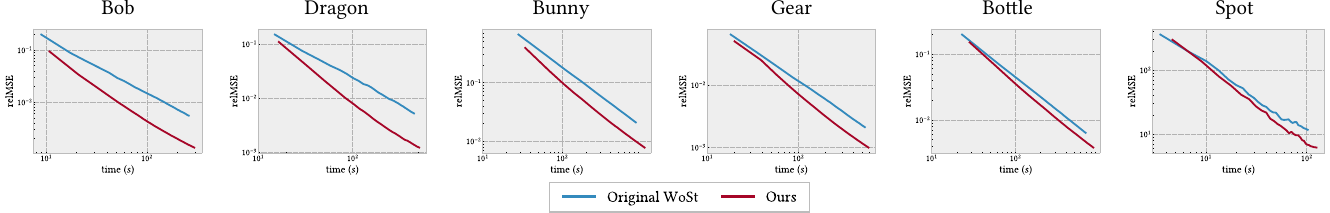}
    \centering
    \caption{relMSE plotted as a function of time of our 3D experiments. Despite the performance overhead, our method still maintains an advantage in equal-time comparisons across all 3D problems.}
    \label{fig:3d_time}
\end{figure*}

Qualitative and quantitative results are presented in \cref{fig:exp_1} and \cref{fig:exp_1_equal_time}, while \cref{fig:3d_wpp} and \cref{fig:3d_time} illustrate the relMSE as a function of wpp and as a function of time, \textit{resp.} The results show that our method outperforms the original WoSt both qualitatively and quantitatively in all problems. Although our method incurs some performance costs, requiring longer runtimes at the same wpp, the results under equal time (\cref{fig:3d_time}) demonstrate that the variance reduction effect of our method compensates for the runtime overhead.

\subsubsection{2D Diffusion Curves}

We use the diffusion curves~\cite{orzan2008dc} as our 2D dataset, which employs Dirichlet boundaries to fill colors in vector graphics. Neumann boundaries are added at the bounding box to form closed domains. We select two representative problems: \emph{Fille} and \emph{Ladybug}. These problems exhibit distinct characteristics: \emph{Fille} has more pronounced light--dark variations; while \emph{Ladybug} features smoother variations, providing a problem more favorable for the baseline.

\begin{figure*}
    \includegraphics[width=0.95\linewidth]{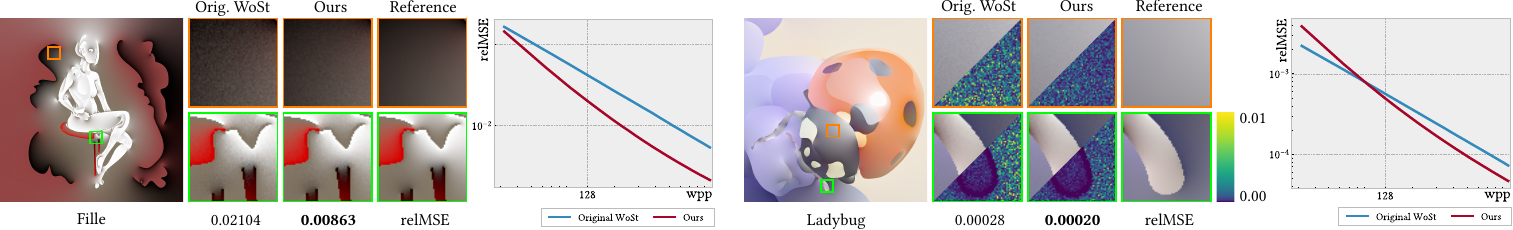}
    \centering
    \caption{Equal-sample qualitative and quantitative results of the 2D diffusion curve experiments. We present the rendering results of two examples at 256 wpp and report their relMSE. Due to the rapid convergence of \textit{Ladybug}, the noise of both methods is nearly imperceptible. To facilitate comparison, we also present false-color maps of the per-pixel relative squared error.}
    \label{fig:diff_curve_quali}
\end{figure*}

\begin{figure*}
    \includegraphics[width=0.95\linewidth]{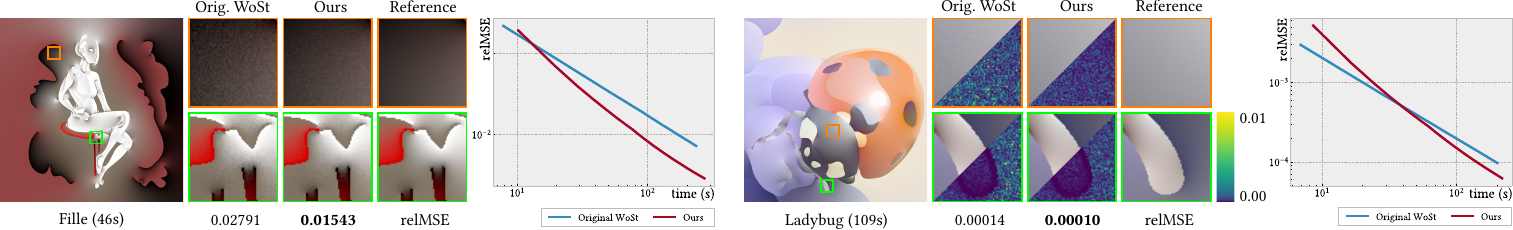}
    \centering
    \caption{Equal-time qualitative and quantitative results of the 2D diffusion curve experiments.}
    \label{fig:diff_curve_quali_et}
\end{figure*}

We report the qualitative and quantitative results in \cref{fig:diff_curve_quali} (equal-sample) and \cref{fig:diff_curve_quali_et} (equal-time).
For the \emph{Ladybug} problem, initially, uniform directional sampling better matches the problem's smoothness compared to the initialization of the guiding distribution. 
The learnable selection probability ensures that, in the early stages of training, unsuitable guiding distributions do not significantly harm the sampling.

\subsection{Evaluation}

\subsubsection{Training Batch Size} 
Our method uses a fraction of batches for training (see \cref{sec:network_design}). To assess the impact of training batch size, we vary the training batch sizes to 64, 128, 256, and 512 wpp, and compare their relMSE and runtime at 1024 wpp. The rest of the experimental setup follows \cref{sec:comp_wost}.
\cref{fig:eval_batch} indicates that the training steps have a significant impact on runtime, primarily because training steps are blocking (\cref{fig:wavefront}). Besides, our model converges relatively quickly, with the performance improvement from 128 wpp to 256 wpp being smaller than that from 64 wpp to 128 wpp. Therefore, using 256 wpp as the training batch size proves to be a reasonable choice.

\subsubsection{Runtime Breakdown}
\begin{figure}[htbp]
    \includegraphics[width=0.9\linewidth]{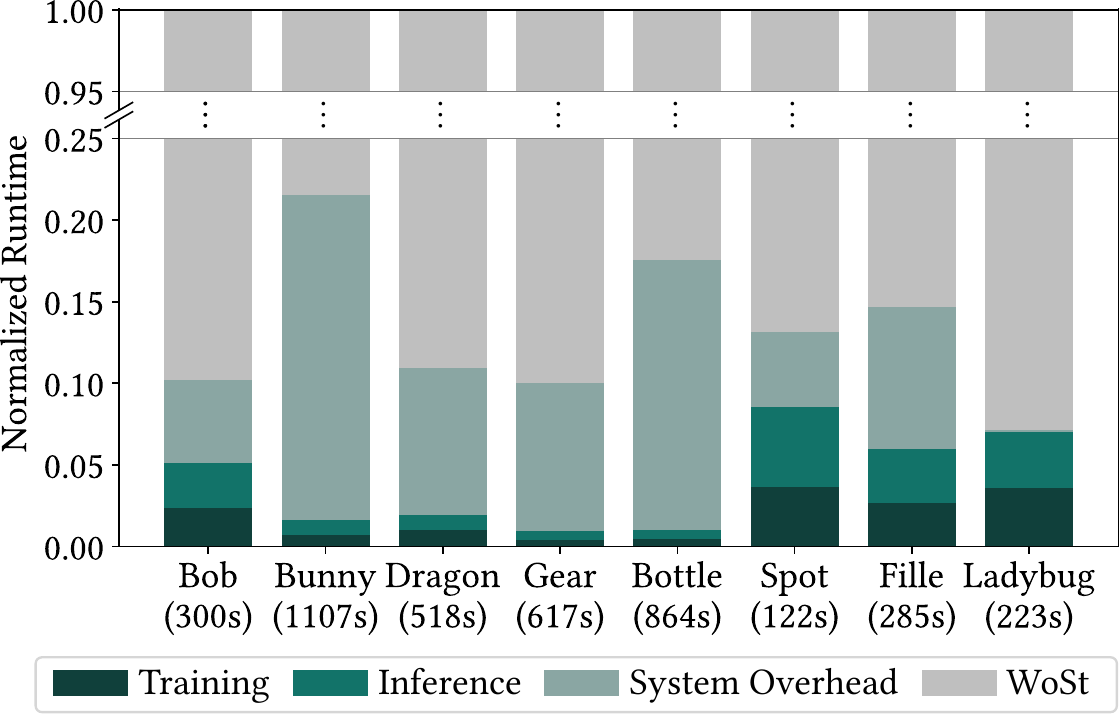}
    \centering
    \caption{Runtime breakdown. The runtime of each execution stage is normalized by the total execution time. \emph{Training} and \emph{Inference} refer to the total time spent on the forward and backward passes of the neural network, \emph{resp.} \emph{System Overhead} represents the additional time spent due to synchronization operations and data structure manipulations introduced by \emph{integrating} the neural network. \emph{WoSt} denotes the time consumed by the original logic of WoSt. On average, our method incurs a 13\% runtime overhead.}
    \label{fig:runtime_overhead}
\end{figure}

We measure the time consumption of training, inference, system overhead, and the original WoSt logic across all problems at 1024 wpp. The results are shown in \cref{fig:runtime_overhead}. The results demonstrate that our method incurs low time overhead, and has substantial room for further system-level optimization.

\subsection{Ablation}
\label{sec:ablation}

\subsubsection{Number of vMF Components in the Mixture Model} 
\label{sec:abl_vmf}
\begin{figure*} 
    \centering
    \begin{minipage}{0.475\textwidth}
        \centering
    \includegraphics[width=0.65\columnwidth]{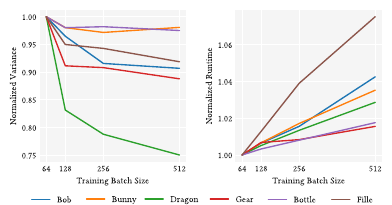}
    \centering
    \caption{The evaluation experiment results for training batch size. We normalize the results by the data with the batch size of 64.}
    \label{fig:eval_batch}
    \end{minipage}
    \hfill
    \begin{minipage}{0.475\textwidth}
        \centering
        \includegraphics[width=0.65\columnwidth]{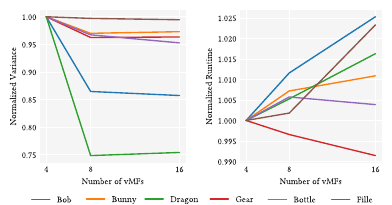}
    \centering
    \caption{The ablation experiment results for the number of vMFs. We normalize the results by the data with the number of 4.}
    \label{fig:eval_vmf}
    \end{minipage}
\end{figure*}

We set the number of vMF components $K$, to 4, 8, and 16, and compare their relMSE and runtime at 1024 wpp. As shown in \cref{fig:eval_vmf}, increasing the number of vMF components from 4 to 8 results in a significant improvement. However, further increasing K to 16 does not provide noticeable benefits and often leads to increased runtime.

\subsubsection{Sample Reflection at Neumann Boundaries}
\label{sec:abl_refl}

We qualitatively evaluate results with or without reflecting samples at Neumann boundaries at 1024 wpp in \cref{fig:abl_refl}. The experiment shows that the method without reflection results in worse variance around the Neumann boundary than the original WoSt, while introducing reflection significantly reduces the variance.

\subsubsection{Multiple Importance Sampling (MIS) with Learnable Selection Probabilities}
\label{sec:abl_mis}

We conduct experiments on problems in \cref{sec:comp_wost}, with the same setup except for different sampling configurations: uniform directional sampling only (selection probabilities $c(x) \equiv 0$), fixed MIS selection probability ($c(x) \equiv 0.5$), guided sampling only ($c(x) \equiv 1$), and our learnable MIS ($c(x)$ is decoded from the neural field). We report the results in \cref{tab:mis}. The results show that our learnable MIS strategy achieves the best performance.

\begin{table}[htbp]
\caption{Quantitative results (relMSE$\downarrow$) of the ablation study on MIS with learnable selection probabilities.}
\label{tab:mis}
\centering
\resizebox{0.9\columnwidth}{!}{%
\begin{tabular}{lcccc}
\toprule
Problem & Uniform Only & Fixed $c(x) \equiv 0.5$ & Guiding Only & Ours \\
\midrule
Bob & 0.00385 & \tbest{0.00181} & \sbest{0.00136} & \best{0.00098} \\
Bunny & 0.01511 & \tbest{0.00690} & \sbest{0.00591} & \best{0.00564} \\
Dragon & 0.00353 & \tbest{0.00158} & \sbest{0.00135} & \best{0.00088} \\
Gear & 0.00153 & \tbest{0.00085} & \sbest{0.00075} & \best{0.00072} \\
Bottle & 0.00459 & \tbest{0.00305} & \sbest{0.00288} & \best{0.00274} \\
Fille & 0.00523 & \sbest{0.00210} & \tbest{0.00241} & \best{0.00208} \\
\bottomrule
\end{tabular}%
}
\end{table}

\section{Limitations and Future Work}

\label{sec:limitations}

\begin{figure*} 
    \centering
    \begin{minipage}{0.475\textwidth}
        \centering
        \includegraphics[width=\textwidth]{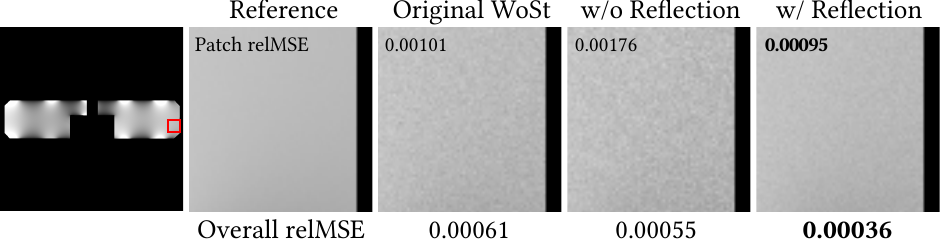}
        \caption{Ablation experiment results for the sample reflection at Neumann boundaries. We show qualitative results of the patch near a Neumann boundary. We also report their patch relMSE and overall relMSE, \emph{resp.}}
        \label{fig:abl_refl}
    \end{minipage}
    \hfill
    \begin{minipage}{0.475\textwidth}
        \centering
        \includegraphics[width=\textwidth]{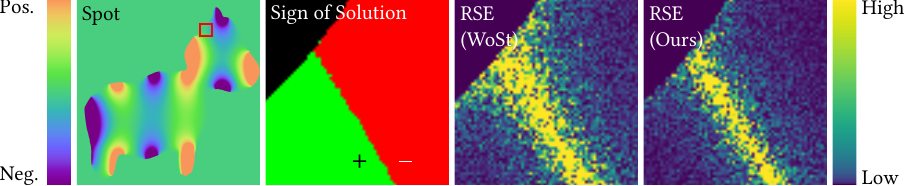}
        \caption{Limitation: Our method effectively reduces shape variance, but like the baseline original WoSt, it cannot reduce sign variance.}
        \label{fig:limitation}
    \end{minipage}
\end{figure*}

\paragraph{Positivization} Our method effectively reduces shape variance; however, for real-valued integrands, we cannot eliminate sign variance with a positive-valued PDF~\cite{art2000safemis}, as shown in \cref{fig:limitation}. While recent differentiable rendering research~\cite{belhe2024brdfderiv, zeltner2021differentialmc, fan2024drpg} has explored similar problems, we believe they are inapplicable to WoSt. Besides, existing Monte Carlo methods for PDE lack consideration of positivization, we therefore regard this issue as beyond our current scope and defer it to future work.

\paragraph{Delta Function} Our method alone cannot handle delta boundaries or source. We anticipate that combining our method with reverse walk splatting~\cite{qi22bidirectional}, or developing an analogy to next event estimation~\cite{kajiya1986re, immel1986radiosity} and integrating it via multiple importance sampling, could address this issue. 

\paragraph{More Estimators} We omit discussions on Robin boundaries~\cite{miller2024robin}, Kelvin-transformed domains~\cite{nabizadeh2021kelvin} and other PDEs~\cite{sawhney2022spatiallyvarying}. However, we believe that extending our method to these cases should not pose significant challenges. Additionally, there should also be a guiding method for \emph{walk on boundary}~\cite{sugimoto2023wob}. For the inverse PDE estimators, drawing from differentiable rendering, it is also expected to require different guiding strategies~\cite{fan2024drpg}.

\paragraph{Importance Sampling Other Terms} Our method performs importance sampling on the recursive term. For the Neumann contribution term, there is already efficient sampling method based on SNCH trees~\cite{sawhney2023wost}. For the source contribution term, using a certain form of light trees~\cite{lin2020lightcuts, conty2018lighttree} may be a good choice. Based on our framework, we may also guide the source sampling using a directional vMF distribution and a radial Beta distribution.

\paragraph{Combination with Other Variance Reduction Methods} Our method is theoretically orthogonal to existing methods. Integrating our method with existing approaches holds the potential to further reduce the variance of WoSt.

\section{Conclusion}

We propose a guiding-based method to importance sample the recursive term of the walk on stars (WoSt) estimator, drawing connections between Monte Carlo PDE solvers and Monte Carlo rendering.
We approximate the directional distribution of the recursive term of WoSt by fitting a guiding distribution, parameterized by a neural field, to observations from previous walks. This distribution is then used to guide subsequent walks with improved sampling efficiency.
A sample reflection technique is introduced to better shape the guiding distribution to align with the target distribution at Neumann boundaries.
Additionally, the learnable selection probabilities adapts the balance between the uniform directional sampling and guiding to the local properties of the solution field, further reducing variance.
Experiments under equal time and equal sample settings show that our method effectively reduces the variance with relatively low overhead.
The expressive guiding distribution improves sampling efficiency, particularly at positions where the solution field exhibits higher frequency.
We anticipate further advancements in variance reduction for Monte Carlo PDE solvers and believe rendering techniques will have broader applications across other fields.

\begin{acks}
We thank anonymous reviewers for their valuable feedback. The \textit{Bob} model is provided by Keenan Crane; the \textit{Dragon} and \textit{Bunny} models are from the Stanford Computer Graphics Laboratory; the \textit{Gear} model is courtesy of \citet{hu2018tetwild}; and the \textit{Spot} model is from \citet{crane2013robust}. \textit{Fille} and \textit{Ladybug} are from \citet{orzan2008dc}. This project is supported by Undergraduate Disruptive Innovation Talent Cultivation Program of Tsinghua University. Tianyu Huang would like to thank Honghao Dong for valuable discussions, as well as Toshiya Hachisuka, Ryusuke Sugimoto, Li Yi, and EECG for their support.
\end{acks}

\bibliographystyle{ACM-Reference-Format}
\bibliography{path-guiding-pde-citations}


\begin{thebibliography}{70}


\ifx \showCODEN    \undefined \def \showCODEN     #1{\unskip}     \fi
\ifx \showDOI      \undefined \def \showDOI       #1{#1}\fi
\ifx \showISBNx    \undefined \def \showISBNx     #1{\unskip}     \fi
\ifx \showISBNxiii \undefined \def \showISBNxiii  #1{\unskip}     \fi
\ifx \showISSN     \undefined \def \showISSN      #1{\unskip}     \fi
\ifx \showLCCN     \undefined \def \showLCCN      #1{\unskip}     \fi
\ifx \shownote     \undefined \def \shownote      #1{#1}          \fi
\ifx \showarticletitle \undefined \def \showarticletitle #1{#1}   \fi
\ifx \showURL      \undefined \def \showURL       {\relax}        \fi
\providecommand\bibfield[2]{#2}
\providecommand\bibinfo[2]{#2}
\providecommand\natexlab[1]{#1}
\providecommand\showeprint[2][]{arXiv:#2}

\bibitem[Bakbouk and Peers(2023)]%
        {bakbouk2023mvc}
\bibfield{author}{\bibinfo{person}{Ghada Bakbouk} {and} \bibinfo{person}{Pieter Peers}.} \bibinfo{year}{2023}\natexlab{}.
\newblock \showarticletitle{{Mean Value Caching for Walk on Spheres}}. In \bibinfo{booktitle}{\emph{Eurographics Symposium on Rendering}}, \bibfield{editor}{\bibinfo{person}{Tobias Ritschel} {and} \bibinfo{person}{Andrea Weidlich}} (Eds.). \bibinfo{publisher}{The Eurographics Association}.
\newblock
\showISBNx{978-3-03868-229-5978-3-03868-228-8}
\showISSN{1727-3463}
\urldef\tempurl%
\url{https://doi.org/10.2312/sr.20231120}
\showDOI{\tempurl}


\bibitem[Bati et~al\mbox{.}(2023)]%
        {bati2023infrared}
\bibfield{author}{\bibinfo{person}{M\'{e}gane Bati}, \bibinfo{person}{St\'{e}phane Blanco}, \bibinfo{person}{Christophe Coustet}, \bibinfo{person}{Vincent Eymet}, \bibinfo{person}{Vincent Forest}, \bibinfo{person}{Richard Fournier}, \bibinfo{person}{Jacques Gautrais}, \bibinfo{person}{Nicolas Mellado}, \bibinfo{person}{Mathias Paulin}, {and} \bibinfo{person}{Benjamin Piaud}.} \bibinfo{year}{2023}\natexlab{}.
\newblock \showarticletitle{Coupling Conduction, Convection and Radiative Transfer in a Single Path-Space: Application to Infrared Rendering}.
\newblock \bibinfo{journal}{\emph{ACM Trans. Graph.}} \bibinfo{volume}{42}, \bibinfo{number}{4}, Article \bibinfo{articleno}{79} (\bibinfo{date}{July} \bibinfo{year}{2023}), \bibinfo{numpages}{20}~pages.
\newblock
\showISSN{0730-0301}
\urldef\tempurl%
\url{https://doi.org/10.1145/3592121}
\showDOI{\tempurl}


\bibitem[Belhe et~al\mbox{.}(2024)]%
        {belhe2024brdfderiv}
\bibfield{author}{\bibinfo{person}{Yash Belhe}, \bibinfo{person}{Bing Xu}, \bibinfo{person}{Sai~Praveen Bangaru}, \bibinfo{person}{Ravi Ramamoorthi}, {and} \bibinfo{person}{Tzu-Mao Li}.} \bibinfo{year}{2024}\natexlab{}.
\newblock \showarticletitle{Importance Sampling BRDF Derivatives}.
\newblock \bibinfo{journal}{\emph{ACM Trans. Graph.}} \bibinfo{volume}{43}, \bibinfo{number}{3}, Article \bibinfo{articleno}{25} (\bibinfo{date}{April} \bibinfo{year}{2024}), \bibinfo{numpages}{21}~pages.
\newblock
\showISSN{0730-0301}
\urldef\tempurl%
\url{https://doi.org/10.1145/3648611}
\showDOI{\tempurl}


\bibitem[Best and Fisher(1979)]%
        {best1979vmf}
\bibfield{author}{\bibinfo{person}{Donald Best} {and} \bibinfo{person}{Nicholas Fisher}.} \bibinfo{year}{1979}\natexlab{}.
\newblock \showarticletitle{Efficient Simulation of the von Mises Distribution}.
\newblock \bibinfo{journal}{\emph{Journal of the Royal Statistical Society. Series C. Applied Statistics}}  \bibinfo{volume}{28} (\bibinfo{date}{01} \bibinfo{year}{1979}).
\newblock
\urldef\tempurl%
\url{https://doi.org/10.2307/2346732}
\showDOI{\tempurl}


\bibitem[Conty~Estevez and Kulla(2018)]%
        {conty2018lighttree}
\bibfield{author}{\bibinfo{person}{Alejandro Conty~Estevez} {and} \bibinfo{person}{Christopher Kulla}.} \bibinfo{year}{2018}\natexlab{}.
\newblock \showarticletitle{Importance Sampling of Many Lights with Adaptive Tree Splitting}.
\newblock \bibinfo{journal}{\emph{Proc. ACM Comput. Graph. Interact. Tech.}} \bibinfo{volume}{1}, \bibinfo{number}{2}, Article \bibinfo{articleno}{25} (\bibinfo{date}{Aug.} \bibinfo{year}{2018}), \bibinfo{numpages}{17}~pages.
\newblock
\urldef\tempurl%
\url{https://doi.org/10.1145/3233305}
\showDOI{\tempurl}


\bibitem[Crane et~al\mbox{.}(2013)]%
        {crane2013robust}
\bibfield{author}{\bibinfo{person}{Keenan Crane}, \bibinfo{person}{Ulrich Pinkall}, {and} \bibinfo{person}{Peter Schr{\"o}der}.} \bibinfo{year}{2013}\natexlab{}.
\newblock \showarticletitle{Robust fairing via conformal curvature flow}.
\newblock \bibinfo{journal}{\emph{ACM Transactions on Graphics (TOG)}} \bibinfo{volume}{32}, \bibinfo{number}{4} (\bibinfo{year}{2013}), \bibinfo{pages}{1--10}.
\newblock


\bibitem[de~Goes and Desbrun(2024)]%
        {degoes2024barycentric}
\bibfield{author}{\bibinfo{person}{Fernando de Goes} {and} \bibinfo{person}{Mathieu Desbrun}.} \bibinfo{year}{2024}\natexlab{}.
\newblock \showarticletitle{Stochastic Computation of Barycentric Coordinates}.
\newblock \bibinfo{journal}{\emph{ACM Trans. Graph.}} \bibinfo{volume}{43}, \bibinfo{number}{4}, Article \bibinfo{articleno}{42} (\bibinfo{date}{July} \bibinfo{year}{2024}), \bibinfo{numpages}{13}~pages.
\newblock
\showISSN{0730-0301}
\urldef\tempurl%
\url{https://doi.org/10.1145/3658131}
\showDOI{\tempurl}


\bibitem[De~Lambilly et~al\mbox{.}(2023)]%
        {delambilly2023heat}
\bibfield{author}{\bibinfo{person}{Auguste De~Lambilly}, \bibinfo{person}{Gabriel Benedetti}, \bibinfo{person}{Nour Rizk}, \bibinfo{person}{Chen Hanqi}, \bibinfo{person}{Siyuan Huang}, \bibinfo{person}{Junnan Qiu}, \bibinfo{person}{David Louapre}, \bibinfo{person}{Raphael Granier De~Cassagnac}, {and} \bibinfo{person}{Damien Rohmer}.} \bibinfo{year}{2023}\natexlab{}.
\newblock \showarticletitle{Heat Simulation on Meshless Crafted-Made Shapes}. In \bibinfo{booktitle}{\emph{Proceedings of the 16th ACM SIGGRAPH Conference on Motion, Interaction and Games}} (Rennes, France) \emph{(\bibinfo{series}{MIG '23})}. \bibinfo{publisher}{Association for Computing Machinery}, \bibinfo{address}{New York, NY, USA}, Article \bibinfo{articleno}{9}, \bibinfo{numpages}{7}~pages.
\newblock
\showISBNx{9798400703935}
\urldef\tempurl%
\url{https://doi.org/10.1145/3623264.3624457}
\showDOI{\tempurl}


\bibitem[Diolatzis et~al\mbox{.}(2020)]%
        {diolatzis2020practical}
\bibfield{author}{\bibinfo{person}{Stavros Diolatzis}, \bibinfo{person}{Adrien Gruson}, \bibinfo{person}{Wenzel Jakob}, \bibinfo{person}{Derek Nowrouzezahrai}, {and} \bibinfo{person}{George Drettakis}.} \bibinfo{year}{2020}\natexlab{}.
\newblock \showarticletitle{Practical Product Path Guiding Using Linearly Transformed Cosines}.
\newblock \bibinfo{journal}{\emph{In Computer Graphics Forum (Proceedings of Eurographics Symposium on Rendering)}} \bibinfo{volume}{39}, \bibinfo{number}{4} (\bibinfo{date}{July} \bibinfo{year}{2020}).
\newblock


\bibitem[Dodik et~al\mbox{.}(2022)]%
        {dorik2022sdmm}
\bibfield{author}{\bibinfo{person}{Ana Dodik}, \bibinfo{person}{Marios Papas}, \bibinfo{person}{Cengiz Öztireli}, {and} \bibinfo{person}{Thomas Müller}.} \bibinfo{year}{2022}\natexlab{}.
\newblock \showarticletitle{Path Guiding Using Spatio-Directional Mixture Models}.
\newblock \bibinfo{journal}{\emph{Computer Graphics Forum}} \bibinfo{volume}{41}, \bibinfo{number}{1} (\bibinfo{year}{2022}), \bibinfo{pages}{172--189}.
\newblock
\urldef\tempurl%
\url{https://doi.org/10.1111/cgf.14428}
\showDOI{\tempurl}
\showeprint{https://onlinelibrary.wiley.com/doi/pdf/10.1111/cgf.14428}


\bibitem[Dong et~al\mbox{.}(2024)]%
        {dong2024efficient}
\bibfield{author}{\bibinfo{person}{Honghao Dong}, \bibinfo{person}{Rui Su}, \bibinfo{person}{Guoping Wang}, {and} \bibinfo{person}{Sheng Li}.} \bibinfo{year}{2024}\natexlab{}.
\newblock \showarticletitle{Efficient Neural Path Guiding with 4D Modeling}. In \bibinfo{booktitle}{\emph{SIGGRAPH Asia 2024 Conference Papers}} \emph{(\bibinfo{series}{SA '24})}. \bibinfo{publisher}{Association for Computing Machinery}, \bibinfo{address}{New York, NY, USA}, Article \bibinfo{articleno}{21}, \bibinfo{numpages}{11}~pages.
\newblock
\showISBNx{9798400711312}
\urldef\tempurl%
\url{https://doi.org/10.1145/3680528.3687687}
\showDOI{\tempurl}


\bibitem[Dong et~al\mbox{.}(2023)]%
        {dong2023npm}
\bibfield{author}{\bibinfo{person}{Honghao Dong}, \bibinfo{person}{Guoping Wang}, {and} \bibinfo{person}{Sheng Li}.} \bibinfo{year}{2023}\natexlab{}.
\newblock \showarticletitle{Neural Parametric Mixtures for Path Guiding}. In \bibinfo{booktitle}{\emph{ACM SIGGRAPH 2023 Conference Proceedings}} (Los Angeles, CA, USA) \emph{(\bibinfo{series}{SIGGRAPH '23})}. \bibinfo{publisher}{Association for Computing Machinery}, \bibinfo{address}{New York, NY, USA}, Article \bibinfo{articleno}{29}, \bibinfo{numpages}{10}~pages.
\newblock
\showISBNx{9798400701597}
\urldef\tempurl%
\url{https://doi.org/10.1145/3588432.3591533}
\showDOI{\tempurl}


\bibitem[Ermakov and Sipin(2009)]%
        {ermakov2009woh}
\bibfield{author}{\bibinfo{person}{S. Ermakov} {and} \bibinfo{person}{A. Sipin}.} \bibinfo{year}{2009}\natexlab{}.
\newblock \showarticletitle{The “walk in hemispheres” process and its applications to solving boundary value problems}.
\newblock \bibinfo{journal}{\emph{Vestnik St. Petersburg University: Mathematics}}  \bibinfo{volume}{42} (\bibinfo{date}{09} \bibinfo{year}{2009}), \bibinfo{pages}{155--163}.
\newblock
\urldef\tempurl%
\url{https://doi.org/10.3103/S1063454109030029}
\showDOI{\tempurl}


\bibitem[Fan et~al\mbox{.}(2023)]%
        {fan2023pgspecular}
\bibfield{author}{\bibinfo{person}{Zhimin Fan}, \bibinfo{person}{Pengpei Hong}, \bibinfo{person}{Jie Guo}, \bibinfo{person}{Changqing Zou}, \bibinfo{person}{Yanwen Guo}, {and} \bibinfo{person}{Ling-Qi Yan}.} \bibinfo{year}{2023}\natexlab{}.
\newblock \showarticletitle{Manifold Path Guiding for Importance Sampling Specular Chains}.
\newblock \bibinfo{journal}{\emph{ACM Trans. Graph.}} \bibinfo{volume}{42}, \bibinfo{number}{6}, Article \bibinfo{articleno}{257} (\bibinfo{date}{Dec.} \bibinfo{year}{2023}), \bibinfo{numpages}{14}~pages.
\newblock
\showISSN{0730-0301}
\urldef\tempurl%
\url{https://doi.org/10.1145/3618360}
\showDOI{\tempurl}


\bibitem[Fan et~al\mbox{.}(2024)]%
        {fan2024drpg}
\bibfield{author}{\bibinfo{person}{Zhimin Fan}, \bibinfo{person}{Pengcheng Shi}, \bibinfo{person}{Mufan Guo}, \bibinfo{person}{Ruoyu Fu}, \bibinfo{person}{Yanwen Guo}, {and} \bibinfo{person}{Jie Guo}.} \bibinfo{year}{2024}\natexlab{}.
\newblock \showarticletitle{Conditional Mixture Path Guiding for Differentiable Rendering}.
\newblock \bibinfo{journal}{\emph{ACM Trans. Graph.}} \bibinfo{volume}{43}, \bibinfo{number}{4}, Article \bibinfo{articleno}{48} (\bibinfo{date}{July} \bibinfo{year}{2024}), \bibinfo{numpages}{11}~pages.
\newblock
\showISSN{0730-0301}
\urldef\tempurl%
\url{https://doi.org/10.1145/3658133}
\showDOI{\tempurl}


\bibitem[Hadadan et~al\mbox{.}(2021)]%
        {hadadan2021nerad}
\bibfield{author}{\bibinfo{person}{Saeed Hadadan}, \bibinfo{person}{Shuhong Chen}, {and} \bibinfo{person}{Matthias Zwicker}.} \bibinfo{year}{2021}\natexlab{}.
\newblock \showarticletitle{Neural radiosity}.
\newblock \bibinfo{journal}{\emph{{ACM} Transactions on Graphics}} \bibinfo{volume}{40}, \bibinfo{number}{6} (\bibinfo{date}{dec} \bibinfo{year}{2021}), \bibinfo{pages}{1--11}.
\newblock
\urldef\tempurl%
\url{https://doi.org/10.1145/3478513.3480569}
\showDOI{\tempurl}


\bibitem[He et~al\mbox{.}(2018)]%
        {he2018slang}
\bibfield{author}{\bibinfo{person}{Yong He}, \bibinfo{person}{Kayvon Fatahalian}, {and} \bibinfo{person}{Tim Foley}.} \bibinfo{year}{2018}\natexlab{}.
\newblock \showarticletitle{Slang: language mechanisms for extensible real-time shading systems}.
\newblock \bibinfo{journal}{\emph{ACM Trans. Graph.}} \bibinfo{volume}{37}, \bibinfo{number}{4}, Article \bibinfo{articleno}{141} (\bibinfo{date}{July} \bibinfo{year}{2018}), \bibinfo{numpages}{13}~pages.
\newblock
\showISSN{0730-0301}
\urldef\tempurl%
\url{https://doi.org/10.1145/3197517.3201380}
\showDOI{\tempurl}


\bibitem[Herholz et~al\mbox{.}(2019)]%
        {herholtz2019vpg}
\bibfield{author}{\bibinfo{person}{Sebastian Herholz}, \bibinfo{person}{Yangyang Zhao}, \bibinfo{person}{Oskar Elek}, \bibinfo{person}{Derek Nowrouzezahrai}, \bibinfo{person}{Hendrik P.~A. Lensch}, {and} \bibinfo{person}{Jaroslav K\v{r}iv\'{a}nek}.} \bibinfo{year}{2019}\natexlab{}.
\newblock \showarticletitle{Volume Path Guiding Based on Zero-Variance Random Walk Theory}.
\newblock \bibinfo{journal}{\emph{ACM Trans. Graph.}} \bibinfo{volume}{38}, \bibinfo{number}{3}, Article \bibinfo{articleno}{25} (\bibinfo{date}{June} \bibinfo{year}{2019}), \bibinfo{numpages}{19}~pages.
\newblock
\showISSN{0730-0301}
\urldef\tempurl%
\url{https://doi.org/10.1145/3230635}
\showDOI{\tempurl}


\bibitem[Hey and Purgathofer(2002)]%
        {hey2002pg}
\bibfield{author}{\bibinfo{person}{Heinrich Hey} {and} \bibinfo{person}{Werner Purgathofer}.} \bibinfo{year}{2002}\natexlab{}.
\newblock \showarticletitle{Importance sampling with hemispherical particle footprints}. In \bibinfo{booktitle}{\emph{Proceedings of the 18th Spring Conference on Computer Graphics}} (Budmerice, Slovakia) \emph{(\bibinfo{series}{SCCG '02})}. \bibinfo{publisher}{Association for Computing Machinery}, \bibinfo{address}{New York, NY, USA}, \bibinfo{pages}{107–114}.
\newblock
\showISBNx{1581136080}
\urldef\tempurl%
\url{https://doi.org/10.1145/584458.584476}
\showDOI{\tempurl}


\bibitem[Hu et~al\mbox{.}(2018)]%
        {hu2018tetwild}
\bibfield{author}{\bibinfo{person}{Yixin Hu}, \bibinfo{person}{Qingnan Zhou}, \bibinfo{person}{Xifeng Gao}, \bibinfo{person}{Alec Jacobson}, \bibinfo{person}{Denis Zorin}, {and} \bibinfo{person}{Daniele Panozzo}.} \bibinfo{year}{2018}\natexlab{}.
\newblock \showarticletitle{Tetrahedral Meshing in the Wild}.
\newblock \bibinfo{journal}{\emph{ACM Trans. Graph.}} \bibinfo{volume}{37}, \bibinfo{number}{4}, Article \bibinfo{articleno}{60} (\bibinfo{date}{July} \bibinfo{year}{2018}), \bibinfo{numpages}{14}~pages.
\newblock
\showISSN{0730-0301}
\urldef\tempurl%
\url{https://doi.org/10.1145/3197517.3201353}
\showDOI{\tempurl}


\bibitem[Huang et~al\mbox{.}(2024)]%
        {huang2024agpg}
\bibfield{author}{\bibinfo{person}{Jiawei Huang}, \bibinfo{person}{Akito Iizuka}, \bibinfo{person}{Hajime Tanaka}, \bibinfo{person}{Taku Komura}, {and} \bibinfo{person}{Yoshifumi Kitamura}.} \bibinfo{year}{2024}\natexlab{}.
\newblock \showarticletitle{Online Neural Path Guiding with Normalized Anisotropic Spherical Gaussians}.
\newblock \bibinfo{journal}{\emph{ACM Trans. Graph.}} \bibinfo{volume}{43}, \bibinfo{number}{3}, Article \bibinfo{articleno}{26} (\bibinfo{date}{April} \bibinfo{year}{2024}), \bibinfo{numpages}{18}~pages.
\newblock
\showISSN{0730-0301}
\urldef\tempurl%
\url{https://doi.org/10.1145/3649310}
\showDOI{\tempurl}


\bibitem[Huo et~al\mbox{.}(2020)]%
        {huo2020conv}
\bibfield{author}{\bibinfo{person}{Yuchi Huo}, \bibinfo{person}{Rui Wang}, \bibinfo{person}{Ruzahng Zheng}, \bibinfo{person}{Hualin Xu}, \bibinfo{person}{Hujun Bao}, {and} \bibinfo{person}{Sung-Eui Yoon}.} \bibinfo{year}{2020}\natexlab{}.
\newblock \showarticletitle{Adaptive Incident Radiance Field Sampling and Reconstruction Using Deep Reinforcement Learning}.
\newblock \bibinfo{journal}{\emph{ACM Trans. Graph.}} \bibinfo{volume}{39}, \bibinfo{number}{1}, Article \bibinfo{articleno}{6} (\bibinfo{date}{Jan.} \bibinfo{year}{2020}), \bibinfo{numpages}{17}~pages.
\newblock
\showISSN{0730-0301}
\urldef\tempurl%
\url{https://doi.org/10.1145/3368313}
\showDOI{\tempurl}


\bibitem[Immel et~al\mbox{.}(1986)]%
        {immel1986radiosity}
\bibfield{author}{\bibinfo{person}{David~S. Immel}, \bibinfo{person}{Michael~F. Cohen}, {and} \bibinfo{person}{Donald~P. Greenberg}.} \bibinfo{year}{1986}\natexlab{}.
\newblock \showarticletitle{A radiosity method for non-diffuse environments}. In \bibinfo{booktitle}{\emph{Proceedings of the 13th Annual Conference on Computer Graphics and Interactive Techniques}} \emph{(\bibinfo{series}{SIGGRAPH '86})}. \bibinfo{publisher}{Association for Computing Machinery}, \bibinfo{address}{New York, NY, USA}, \bibinfo{pages}{133–142}.
\newblock
\showISBNx{0897911962}
\urldef\tempurl%
\url{https://doi.org/10.1145/15922.15901}
\showDOI{\tempurl}


\bibitem[Jain et~al\mbox{.}(2024)]%
        {jain2024mcfluid}
\bibfield{author}{\bibinfo{person}{Pranav Jain}, \bibinfo{person}{Ziyin Qu}, \bibinfo{person}{Peter~Yichen Chen}, {and} \bibinfo{person}{Oded Stein}.} \bibinfo{year}{2024}\natexlab{}.
\newblock \showarticletitle{Neural Monte Carlo Fluid Simulation}. In \bibinfo{booktitle}{\emph{ACM SIGGRAPH 2024 Conference Papers}} (Denver, CO, USA) \emph{(\bibinfo{series}{SIGGRAPH '24})}. \bibinfo{publisher}{Association for Computing Machinery}, \bibinfo{address}{New York, NY, USA}, Article \bibinfo{articleno}{9}, \bibinfo{numpages}{11}~pages.
\newblock
\showISBNx{9798400705250}
\urldef\tempurl%
\url{https://doi.org/10.1145/3641519.3657438}
\showDOI{\tempurl}


\bibitem[Jensen(1995)]%
        {jensen1995pm}
\bibfield{author}{\bibinfo{person}{Henrik~Wann Jensen}.} \bibinfo{year}{1995}\natexlab{}.
\newblock \showarticletitle{Importance Driven Path Tracing using the Photon Map}. In \bibinfo{booktitle}{\emph{Rendering Techniques}}.
\newblock
\urldef\tempurl%
\url{https://api.semanticscholar.org/CorpusID:9344202}
\showURL{%
\tempurl}


\bibitem[Kajiya(1986)]%
        {kajiya1986re}
\bibfield{author}{\bibinfo{person}{James~T. Kajiya}.} \bibinfo{year}{1986}\natexlab{}.
\newblock \showarticletitle{The rendering equation}. In \bibinfo{booktitle}{\emph{Proceedings of the 13th Annual Conference on Computer Graphics and Interactive Techniques}} \emph{(\bibinfo{series}{SIGGRAPH '86})}. \bibinfo{publisher}{Association for Computing Machinery}, \bibinfo{address}{New York, NY, USA}, \bibinfo{pages}{143–150}.
\newblock
\showISBNx{0897911962}
\urldef\tempurl%
\url{https://doi.org/10.1145/15922.15902}
\showDOI{\tempurl}


\bibitem[Karras(2012)]%
        {karras2012lbvh}
\bibfield{author}{\bibinfo{person}{Tero Karras}.} \bibinfo{year}{2012}\natexlab{}.
\newblock \showarticletitle{Maximizing parallelism in the construction of BVHs, octrees, and k-d trees}. In \bibinfo{booktitle}{\emph{Proceedings of the Fourth ACM SIGGRAPH / Eurographics Conference on High-Performance Graphics}} (Paris, France) \emph{(\bibinfo{series}{EGGH-HPG'12})}. \bibinfo{publisher}{Eurographics Association}, \bibinfo{address}{Goslar, DEU}, \bibinfo{pages}{33–37}.
\newblock
\showISBNx{9783905674415}


\bibitem[Kingma(2014)]%
        {kingma2014adam}
\bibfield{author}{\bibinfo{person}{Diederik~P Kingma}.} \bibinfo{year}{2014}\natexlab{}.
\newblock \showarticletitle{Adam: A method for stochastic optimization}.
\newblock \bibinfo{journal}{\emph{arXiv preprint arXiv:1412.6980}} (\bibinfo{year}{2014}).
\newblock


\bibitem[Laine et~al\mbox{.}(2013)]%
        {laine2013megakernel}
\bibfield{author}{\bibinfo{person}{Samuli Laine}, \bibinfo{person}{Tero Karras}, {and} \bibinfo{person}{Timo Aila}.} \bibinfo{year}{2013}\natexlab{}.
\newblock \showarticletitle{Megakernels considered harmful: wavefront path tracing on GPUs}. In \bibinfo{booktitle}{\emph{Proceedings of the 5th High-Performance Graphics Conference}} (Anaheim, California) \emph{(\bibinfo{series}{HPG '13})}. \bibinfo{publisher}{Association for Computing Machinery}, \bibinfo{address}{New York, NY, USA}, \bibinfo{pages}{137–143}.
\newblock
\showISBNx{9781450321358}
\urldef\tempurl%
\url{https://doi.org/10.1145/2492045.2492060}
\showDOI{\tempurl}


\bibitem[Li et~al\mbox{.}(2022)]%
        {li2022pathcut}
\bibfield{author}{\bibinfo{person}{He Li}, \bibinfo{person}{Beibei Wang}, \bibinfo{person}{Changehe Tu}, \bibinfo{person}{Kun Xu}, \bibinfo{person}{Nicolas Holzschuch}, {and} \bibinfo{person}{Ling-Qi Yan}.} \bibinfo{year}{2022}\natexlab{}.
\newblock \showarticletitle{Unbiased Caustics Rendering Guided by Representative Specular Paths}. In \bibinfo{booktitle}{\emph{Proceedings of SIGGRAPH Asia 2022}}.
\newblock


\bibitem[Li et~al\mbox{.}(2023)]%
        {li2023neuralcaches}
\bibfield{author}{\bibinfo{person}{Zilu Li}, \bibinfo{person}{Guandao Yang}, \bibinfo{person}{Xi Deng}, \bibinfo{person}{Christopher De~Sa}, \bibinfo{person}{Bharath Hariharan}, {and} \bibinfo{person}{Steve Marschner}.} \bibinfo{year}{2023}\natexlab{}.
\newblock \showarticletitle{Neural Caches for Monte Carlo Partial Differential Equation Solvers}. In \bibinfo{booktitle}{\emph{SIGGRAPH Asia 2023 Conference Papers}} (Sydney, NSW, Australia) \emph{(\bibinfo{series}{SA '23})}. \bibinfo{publisher}{Association for Computing Machinery}, \bibinfo{address}{New York, NY, USA}, Article \bibinfo{articleno}{34}, \bibinfo{numpages}{10}~pages.
\newblock
\showISBNx{9798400703157}
\urldef\tempurl%
\url{https://doi.org/10.1145/3610548.3618141}
\showDOI{\tempurl}


\bibitem[Li et~al\mbox{.}(2024)]%
        {li2024ncv}
\bibfield{author}{\bibinfo{person}{Zilu Li}, \bibinfo{person}{Guandao Yang}, \bibinfo{person}{Qingqing Zhao}, \bibinfo{person}{Xi Deng}, \bibinfo{person}{Leonidas Guibas}, \bibinfo{person}{Bharath Hariharan}, {and} \bibinfo{person}{Gordon Wetzstein}.} \bibinfo{year}{2024}\natexlab{}.
\newblock \showarticletitle{Neural Control Variates with Automatic Integration}. In \bibinfo{booktitle}{\emph{ACM SIGGRAPH 2024 Conference Papers}} (Denver, CO, USA) \emph{(\bibinfo{series}{SIGGRAPH '24})}. \bibinfo{publisher}{Association for Computing Machinery}, \bibinfo{address}{New York, NY, USA}, Article \bibinfo{articleno}{10}, \bibinfo{numpages}{9}~pages.
\newblock
\showISBNx{9798400705250}
\urldef\tempurl%
\url{https://doi.org/10.1145/3641519.3657395}
\showDOI{\tempurl}


\bibitem[Lin and Yuksel(2020)]%
        {lin2020lightcuts}
\bibfield{author}{\bibinfo{person}{Daqi Lin} {and} \bibinfo{person}{Cem Yuksel}.} \bibinfo{year}{2020}\natexlab{}.
\newblock \showarticletitle{Real-Time Stochastic Lightcuts}.
\newblock \bibinfo{journal}{\emph{Proc. ACM Comput. Graph. Interact. Tech. (Proceedings of I3D 2020)}} \bibinfo{volume}{3}, \bibinfo{number}{1} (\bibinfo{year}{2020}), \bibinfo{numpages}{18}~pages.
\newblock
\urldef\tempurl%
\url{https://doi.org/10.1145/3384543}
\showDOI{\tempurl}


\bibitem[Miller et~al\mbox{.}(2023)]%
        {miller2023bvc}
\bibfield{author}{\bibinfo{person}{Bailey Miller}, \bibinfo{person}{Rohan Sawhney}, \bibinfo{person}{Keenan Crane}, {and} \bibinfo{person}{Ioannis Gkioulekas}.} \bibinfo{year}{2023}\natexlab{}.
\newblock \showarticletitle{Boundary Value Caching for Walk on Spheres}.
\newblock \bibinfo{journal}{\emph{ACM Trans. Graph.}} \bibinfo{volume}{42}, \bibinfo{number}{4} (\bibinfo{year}{2023}).
\newblock


\bibitem[Miller et~al\mbox{.}(2024a)]%
        {miller2024differential}
\bibfield{author}{\bibinfo{person}{Bailey Miller}, \bibinfo{person}{Rohan Sawhney}, \bibinfo{person}{Keenan Crane}, {and} \bibinfo{person}{Ioannis Gkioulekas}.} \bibinfo{year}{2024}\natexlab{a}.
\newblock \showarticletitle{Differential Walk on Spheres}.
\newblock \bibinfo{journal}{\emph{ACM Trans. Graph.}} \bibinfo{volume}{43}, \bibinfo{number}{6} (\bibinfo{year}{2024}).
\newblock


\bibitem[Miller et~al\mbox{.}(2024b)]%
        {miller2024robin}
\bibfield{author}{\bibinfo{person}{Bailey Miller}, \bibinfo{person}{Rohan Sawhney}, \bibinfo{person}{Keenan Crane}, {and} \bibinfo{person}{Ioannis Gkioulekas}.} \bibinfo{year}{2024}\natexlab{b}.
\newblock \showarticletitle{Walkin' Robin: Walk on Stars with Robin Boundary Conditions}.
\newblock \bibinfo{journal}{\emph{ACM Trans. Graph.}} \bibinfo{volume}{43}, \bibinfo{number}{4} (\bibinfo{year}{2024}).
\newblock


\bibitem[Muchacho and Pokorny(2024)]%
        {muchacho2024pathplanning}
\bibfield{author}{\bibinfo{person}{Rafael I.~Cabral Muchacho} {and} \bibinfo{person}{Florian~T. Pokorny}.} \bibinfo{year}{2024}\natexlab{}.
\newblock \bibinfo{title}{Walk on Spheres for PDE-based Path Planning}.
\newblock
\newblock
\showeprint[arxiv]{2406.01713}~[cs.RO]
\urldef\tempurl%
\url{https://arxiv.org/abs/2406.01713}
\showURL{%
\tempurl}


\bibitem[Muller(1956)]%
        {muller1956wos}
\bibfield{author}{\bibinfo{person}{Mervin~E. Muller}.} \bibinfo{year}{1956}\natexlab{}.
\newblock \showarticletitle{{Some Continuous Monte Carlo Methods for the Dirichlet Problem}}.
\newblock \bibinfo{journal}{\emph{The Annals of Mathematical Statistics}} \bibinfo{volume}{27}, \bibinfo{number}{3} (\bibinfo{year}{1956}), \bibinfo{pages}{569 -- 589}.
\newblock
\urldef\tempurl%
\url{https://doi.org/10.1214/aoms/1177728169}
\showDOI{\tempurl}


\bibitem[M\"uller(2021)]%
        {tiny-cuda-nn}
\bibfield{author}{\bibinfo{person}{Thomas M\"uller}.} \bibinfo{year}{2021}\natexlab{}.
\newblock \bibinfo{booktitle}{\emph{{tiny-cuda-nn}}}.
\newblock
\urldef\tempurl%
\url{https://github.com/NVlabs/tiny-cuda-nn}
\showURL{%
\tempurl}


\bibitem[M\"uller et~al\mbox{.}(2017)]%
        {mueller2017ppg}
\bibfield{author}{\bibinfo{person}{Thomas M\"uller}, \bibinfo{person}{Markus Gross}, {and} \bibinfo{person}{Jan Nov\'ak}.} \bibinfo{year}{2017}\natexlab{}.
\newblock \showarticletitle{Practical Path Guiding for Efficient Light-Transport Simulation}.
\newblock \bibinfo{journal}{\emph{Computer Graphics Forum (Proceedings of EGSR)}} \bibinfo{volume}{36}, \bibinfo{number}{4} (\bibinfo{date}{June} \bibinfo{year}{2017}), \bibinfo{pages}{91--100}.
\newblock
\urldef\tempurl%
\url{https://doi.org/10.1111/cgf.13227}
\showDOI{\tempurl}


\bibitem[M\"{u}ller et~al\mbox{.}(2019)]%
        {mueller2019neural}
\bibfield{author}{\bibinfo{person}{Thomas M\"{u}ller}, \bibinfo{person}{Brian McWilliams}, \bibinfo{person}{Fabrice Rousselle}, \bibinfo{person}{Markus Gross}, {and} \bibinfo{person}{Jan Nov\'{a}k}.} \bibinfo{year}{2019}\natexlab{}.
\newblock \showarticletitle{Neural Importance Sampling}.
\newblock \bibinfo{journal}{\emph{ACM Trans. Graph.}} \bibinfo{volume}{38}, \bibinfo{number}{5}, Article \bibinfo{articleno}{145} (\bibinfo{date}{Oct.} \bibinfo{year}{2019}), \bibinfo{numpages}{19}~pages.
\newblock
\showISSN{0730-0301}
\urldef\tempurl%
\url{https://doi.org/10.1145/3341156}
\showDOI{\tempurl}


\bibitem[M\"{u}ller et~al\mbox{.}(2021)]%
        {muller2021nrc}
\bibfield{author}{\bibinfo{person}{Thomas M\"{u}ller}, \bibinfo{person}{Fabrice Rousselle}, \bibinfo{person}{Jan Nov\'{a}k}, {and} \bibinfo{person}{Alexander Keller}.} \bibinfo{year}{2021}\natexlab{}.
\newblock \showarticletitle{Real-time neural radiance caching for path tracing}.
\newblock \bibinfo{journal}{\emph{ACM Trans. Graph.}} \bibinfo{volume}{40}, \bibinfo{number}{4}, Article \bibinfo{articleno}{36} (\bibinfo{date}{July} \bibinfo{year}{2021}), \bibinfo{numpages}{16}~pages.
\newblock
\showISSN{0730-0301}
\urldef\tempurl%
\url{https://doi.org/10.1145/3450626.3459812}
\showDOI{\tempurl}


\bibitem[Museth(2021)]%
        {nanovdb}
\bibfield{author}{\bibinfo{person}{Ken Museth}.} \bibinfo{year}{2021}\natexlab{}.
\newblock \showarticletitle{NanoVDB: A GPU-Friendly and Portable VDB Data Structure For Real-Time Rendering And Simulation}. In \bibinfo{booktitle}{\emph{ACM SIGGRAPH 2021 Talks}} (Virtual Event, USA) \emph{(\bibinfo{series}{SIGGRAPH '21})}. \bibinfo{publisher}{Association for Computing Machinery}, \bibinfo{address}{New York, NY, USA}, Article \bibinfo{articleno}{1}, \bibinfo{numpages}{2}~pages.
\newblock
\showISBNx{9781450383738}
\urldef\tempurl%
\url{https://doi.org/10.1145/3450623.3464653}
\showDOI{\tempurl}


\bibitem[Nabizadeh et~al\mbox{.}(2021)]%
        {nabizadeh2021kelvin}
\bibfield{author}{\bibinfo{person}{Mohammad~Sina Nabizadeh}, \bibinfo{person}{Ravi Ramamoorthi}, {and} \bibinfo{person}{Albert Chern}.} \bibinfo{year}{2021}\natexlab{}.
\newblock \showarticletitle{Kelvin transformations for simulations on infinite domains}.
\newblock \bibinfo{journal}{\emph{ACM Transactions on Graphics (TOG)}} \bibinfo{volume}{40}, \bibinfo{number}{4} (\bibinfo{year}{2021}), \bibinfo{pages}{97:1--97:15}.
\newblock


\bibitem[Nam et~al\mbox{.}(2024)]%
        {nam2024solving}
\bibfield{author}{\bibinfo{person}{Hong~Chul Nam}, \bibinfo{person}{Julius Berner}, {and} \bibinfo{person}{Anima Anandkumar}.} \bibinfo{year}{2024}\natexlab{}.
\newblock \showarticletitle{Solving Poisson Equations Using Neural Walk-on-Spheres}. In \bibinfo{booktitle}{\emph{Forty-first International Conference on Machine Learning}}.
\newblock


\bibitem[Orzan et~al\mbox{.}(2008)]%
        {orzan2008dc}
\bibfield{author}{\bibinfo{person}{Alexandrina Orzan}, \bibinfo{person}{Adrien Bousseau}, \bibinfo{person}{Holger Winnem\"{o}ller}, \bibinfo{person}{Pascal Barla}, \bibinfo{person}{Jo\"{e}lle Thollot}, {and} \bibinfo{person}{David Salesin}.} \bibinfo{year}{2008}\natexlab{}.
\newblock \showarticletitle{Diffusion curves: a vector representation for smooth-shaded images}. In \bibinfo{booktitle}{\emph{ACM SIGGRAPH 2008 Papers}} (Los Angeles, California) \emph{(\bibinfo{series}{SIGGRAPH '08})}. \bibinfo{publisher}{Association for Computing Machinery}, \bibinfo{address}{New York, NY, USA}, Article \bibinfo{articleno}{92}, \bibinfo{numpages}{8}~pages.
\newblock
\showISBNx{9781450301121}
\urldef\tempurl%
\url{https://doi.org/10.1145/1399504.1360691}
\showDOI{\tempurl}


\bibitem[Owen and Zhou(2000)]%
        {art2000safemis}
\bibfield{author}{\bibinfo{person}{Art Owen} {and} \bibinfo{person}{Yi Zhou}.} \bibinfo{year}{2000}\natexlab{}.
\newblock \showarticletitle{Safe and Effective Importance Sampling}.
\newblock \bibinfo{journal}{\emph{J. Amer. Statist. Assoc.}} \bibinfo{volume}{95}, \bibinfo{number}{449} (\bibinfo{year}{2000}), \bibinfo{pages}{135--143}.
\newblock
\showISSN{01621459, 1537274X}
\urldef\tempurl%
\url{http://www.jstor.org/stable/2669533}
\showURL{%
\tempurl}


\bibitem[Qi et~al\mbox{.}(2022)]%
        {qi22bidirectional}
\bibfield{author}{\bibinfo{person}{Yang Qi}, \bibinfo{person}{Dario Seyb}, \bibinfo{person}{Benedikt Bitterli}, {and} \bibinfo{person}{Wojciech Jarosz}.} \bibinfo{year}{2022}\natexlab{}.
\newblock \showarticletitle{A bidirectional formulation for {Walk} on {Spheres}}.
\newblock \bibinfo{journal}{\emph{Computer Graphics Forum (Proceedings of EGSR)}} \bibinfo{volume}{41}, \bibinfo{number}{4} (\bibinfo{date}{July} \bibinfo{year}{2022}).
\newblock
\showISSN{1467-8659}
\urldef\tempurl%
\url{https://doi.org/10/jgzr}
\showDOI{\tempurl}


\bibitem[Rath et~al\mbox{.}(2020)]%
        {rath2020vapg}
\bibfield{author}{\bibinfo{person}{Alexander Rath}, \bibinfo{person}{Pascal Grittmann}, \bibinfo{person}{Sebastian Herholz}, \bibinfo{person}{Petr V\'{e}voda}, \bibinfo{person}{Philipp Slusallek}, {and} \bibinfo{person}{Jaroslav K\v{r}iv\'{a}nek}.} \bibinfo{year}{2020}\natexlab{}.
\newblock \showarticletitle{Variance-aware path guiding}.
\newblock \bibinfo{journal}{\emph{ACM Trans. Graph.}} \bibinfo{volume}{39}, \bibinfo{number}{4}, Article \bibinfo{articleno}{151} (\bibinfo{date}{Aug.} \bibinfo{year}{2020}), \bibinfo{numpages}{12}~pages.
\newblock
\showISSN{0730-0301}
\urldef\tempurl%
\url{https://doi.org/10.1145/3386569.3392441}
\showDOI{\tempurl}


\bibitem[Reibold et~al\mbox{.}(2018)]%
        {reibold2018completepathguiding}
\bibfield{author}{\bibinfo{person}{Florian Reibold}, \bibinfo{person}{Johannes Hanika}, \bibinfo{person}{Alisa Jung}, {and} \bibinfo{person}{Carsten Dachsbacher}.} \bibinfo{year}{2018}\natexlab{}.
\newblock \showarticletitle{Selective guided sampling with complete light transport paths}.
\newblock \bibinfo{journal}{\emph{ACM Trans. Graph.}} \bibinfo{volume}{37}, \bibinfo{number}{6}, Article \bibinfo{articleno}{223} (\bibinfo{date}{Dec.} \bibinfo{year}{2018}), \bibinfo{numpages}{14}~pages.
\newblock
\showISSN{0730-0301}
\urldef\tempurl%
\url{https://doi.org/10.1145/3272127.3275030}
\showDOI{\tempurl}


\bibitem[Rioux-Lavoie et~al\mbox{.}(2022)]%
        {lavoie2022mcfluid}
\bibfield{author}{\bibinfo{person}{Damien Rioux-Lavoie}, \bibinfo{person}{Ryusuke Sugimoto}, \bibinfo{person}{Tümay Özdemir}, \bibinfo{person}{Naoharu~H. Shimada}, \bibinfo{person}{Christopher Batty}, \bibinfo{person}{Derek Nowrouzezahrai}, {and} \bibinfo{person}{Toshiya Hachisuka}.} \bibinfo{year}{2022}\natexlab{}.
\newblock \showarticletitle{A Monte Carlo Method for Fluid Simulation}.
\newblock \bibinfo{journal}{\emph{ACM Transactions on Graphics}} \bibinfo{volume}{41}, \bibinfo{number}{6} (\bibinfo{date}{Dec.} \bibinfo{year}{2022}).
\newblock
\urldef\tempurl%
\url{https://doi.org/10.1145/3550454.3555450}
\showDOI{\tempurl}


\bibitem[Ruppert et~al\mbox{.}(2020)]%
        {ruppert2020pavmm}
\bibfield{author}{\bibinfo{person}{Lukas Ruppert}, \bibinfo{person}{Sebastian Herholz}, {and} \bibinfo{person}{Hendrik P.~A. Lensch}.} \bibinfo{year}{2020}\natexlab{}.
\newblock \showarticletitle{Robust fitting of parallax-aware mixtures for path guiding}.
\newblock \bibinfo{journal}{\emph{ACM Trans. Graph.}} \bibinfo{volume}{39}, \bibinfo{number}{4}, Article \bibinfo{articleno}{147} (\bibinfo{date}{Aug.} \bibinfo{year}{2020}), \bibinfo{numpages}{15}~pages.
\newblock
\showISSN{0730-0301}
\urldef\tempurl%
\url{https://doi.org/10.1145/3386569.3392421}
\showDOI{\tempurl}


\bibitem[Sawhney(2021)]%
        {FCPW}
\bibfield{author}{\bibinfo{person}{Rohan Sawhney}.} \bibinfo{year}{2021}\natexlab{}.
\newblock \bibinfo{booktitle}{\emph{FCPW: Fastest Closest Points in the West}}.
\newblock


\bibitem[Sawhney and Crane(2020)]%
        {sawhney2020mcgp}
\bibfield{author}{\bibinfo{person}{Rohan Sawhney} {and} \bibinfo{person}{Keenan Crane}.} \bibinfo{year}{2020}\natexlab{}.
\newblock \showarticletitle{Monte Carlo Geometry Processing: A Grid-Free Approach to PDE-Based Methods on Volumetric Domains}.
\newblock \bibinfo{journal}{\emph{ACM Trans. Graph.}} \bibinfo{volume}{39}, \bibinfo{number}{4} (\bibinfo{year}{2020}).
\newblock


\bibitem[Sawhney and Miller(2023)]%
        {Zombie}
\bibfield{author}{\bibinfo{person}{Rohan Sawhney} {and} \bibinfo{person}{Bailey Miller}.} \bibinfo{year}{2023}\natexlab{}.
\newblock \bibinfo{booktitle}{\emph{{Zombie: A Grid-Free Monte Carlo Solver for PDEs}}}.
\newblock


\bibitem[Sawhney and Miller(2024)]%
        {sawhney2024mcgpcourse}
\bibfield{author}{\bibinfo{person}{Rohan Sawhney} {and} \bibinfo{person}{Bailey Miller}.} \bibinfo{year}{2024}\natexlab{}.
\newblock \showarticletitle{Monte Carlo Geometry Processing}. In \bibinfo{booktitle}{\emph{SGP 2024 Graduate School Courses}}.
\newblock


\bibitem[Sawhney et~al\mbox{.}(2023)]%
        {sawhney2023wost}
\bibfield{author}{\bibinfo{person}{Rohan Sawhney}, \bibinfo{person}{Bailey Miller}, \bibinfo{person}{Ioannis Gkioulekas}, {and} \bibinfo{person}{Keenan Crane}.} \bibinfo{year}{2023}\natexlab{}.
\newblock \showarticletitle{Walk on Stars: A Grid-Free Monte Carlo Method for PDEs with Neumann Boundary Conditions}.
\newblock \bibinfo{journal}{\emph{ACM Trans. Graph.}} \bibinfo{volume}{42}, \bibinfo{number}{4} (\bibinfo{year}{2023}).
\newblock


\bibitem[Sawhney et~al\mbox{.}(2022)]%
        {sawhney2022spatiallyvarying}
\bibfield{author}{\bibinfo{person}{Rohan Sawhney}, \bibinfo{person}{Dario Seyb}, \bibinfo{person}{Wojciech Jarosz}, {and} \bibinfo{person}{Keenan Crane}.} \bibinfo{year}{2022}\natexlab{}.
\newblock \showarticletitle{Grid-free Monte Carlo for PDEs with spatially varying coefficients}.
\newblock \bibinfo{journal}{\emph{ACM Trans. Graph.}} \bibinfo{volume}{41}, \bibinfo{number}{4}, Article \bibinfo{articleno}{53} (\bibinfo{date}{July} \bibinfo{year}{2022}), \bibinfo{numpages}{17}~pages.
\newblock
\showISSN{0730-0301}
\urldef\tempurl%
\url{https://doi.org/10.1145/3528223.3530134}
\showDOI{\tempurl}


\bibitem[Simonov(2008)]%
        {simonov2008wost}
\bibfield{author}{\bibinfo{person}{Nikolai~A. Simonov}.} \bibinfo{year}{2008}\natexlab{}.
\newblock \showarticletitle{Walk-on-Spheres Algorithm for Solving Boundary-Value Problems with Continuity Flux Conditions}.
\newblock
\urldef\tempurl%
\url{https://api.semanticscholar.org/CorpusID:117970575}
\showURL{%
\tempurl}


\bibitem[Sugimoto et~al\mbox{.}(2024a)]%
        {sugimoto2024mcfluid}
\bibfield{author}{\bibinfo{person}{Ryusuke Sugimoto}, \bibinfo{person}{Christopher Batty}, {and} \bibinfo{person}{Toshiya Hachisuka}.} \bibinfo{year}{2024}\natexlab{a}.
\newblock \showarticletitle{Velocity-Based Monte Carlo Fluids}. In \bibinfo{booktitle}{\emph{ACM SIGGRAPH 2024 Conference Papers}} (Denver, CO, USA) \emph{(\bibinfo{series}{SIGGRAPH '24})}. \bibinfo{publisher}{Association for Computing Machinery}, \bibinfo{address}{New York, NY, USA}, Article \bibinfo{articleno}{8}, \bibinfo{numpages}{11}~pages.
\newblock
\showISBNx{9798400705250}
\urldef\tempurl%
\url{https://doi.org/10.1145/3641519.3657405}
\showDOI{\tempurl}


\bibitem[Sugimoto et~al\mbox{.}(2023)]%
        {sugimoto2023wob}
\bibfield{author}{\bibinfo{person}{Ryusuke Sugimoto}, \bibinfo{person}{Terry Chen}, \bibinfo{person}{Yiti Jiang}, \bibinfo{person}{Christopher Batty}, {and} \bibinfo{person}{Toshiya Hachisuka}.} \bibinfo{year}{2023}\natexlab{}.
\newblock \showarticletitle{A Practical Walk-on-Boundary Method for Boundary Value Problems}.
\newblock \bibinfo{journal}{\emph{ACM Trans. Graph.}} \bibinfo{volume}{42}, \bibinfo{number}{4}, Article \bibinfo{articleno}{81} (\bibinfo{date}{jul} \bibinfo{year}{2023}), \bibinfo{numpages}{16}~pages.
\newblock
\urldef\tempurl%
\url{https://doi.org/10.1145/3592109}
\showDOI{\tempurl}


\bibitem[Sugimoto et~al\mbox{.}(2024b)]%
        {sugimoto2024pwos}
\bibfield{author}{\bibinfo{person}{Ryusuke Sugimoto}, \bibinfo{person}{Nathan King}, \bibinfo{person}{Toshiya Hachisuka}, {and} \bibinfo{person}{Christopher Batty}.} \bibinfo{year}{2024}\natexlab{b}.
\newblock \showarticletitle{Projected Walk on Spheres: A Monte Carlo Closest Point Method for Surface PDEs}. In \bibinfo{booktitle}{\emph{ACM SIGGRAPH Asia 2024 Conference Papers}} (Tokyo, Japan) \emph{(\bibinfo{series}{SIGGRAPH Asia '24})}. \bibinfo{publisher}{Association for Computing Machinery}, \bibinfo{address}{New York, NY, USA}, \bibinfo{numpages}{10}~pages.
\newblock
\urldef\tempurl%
\url{https://doi.org/10.1145/3680528.3687599}
\showDOI{\tempurl}


\bibitem[Tokuyoshi(2025)]%
        {tokuyoshi2025vmf}
\bibfield{author}{\bibinfo{person}{Yusuke Tokuyoshi}.} \bibinfo{year}{2025}\natexlab{}.
\newblock \bibinfo{booktitle}{\emph{{A Numerically Stable Implementation of the von Mises–Fisher Distribution on $S^2$}}}.
\newblock \bibinfo{type}{{T}echnical {R}eport}. \bibinfo{institution}{Advanced Micro Devices, Inc.}
\newblock
\urldef\tempurl%
\url{https://gpuopen.com/download/publications/A_Numerically_Stable_Implementation_of_the_von_Mises%E2%80%93Fisher_Distribution_on_S2.pdf}
\showURL{%
\tempurl}


\bibitem[Veach and Guibas(1995a)]%
        {veach1995bdpt}
\bibfield{author}{\bibinfo{person}{Eric Veach} {and} \bibinfo{person}{Leonidas Guibas}.} \bibinfo{year}{1995}\natexlab{a}.
\newblock \showarticletitle{Bidirectional Estimators for Light Transport}. In \bibinfo{booktitle}{\emph{Photorealistic Rendering Techniques}}, \bibfield{editor}{\bibinfo{person}{Georgios Sakas}, \bibinfo{person}{Stefan M{\"u}ller}, {and} \bibinfo{person}{Peter Shirley}} (Eds.). \bibinfo{publisher}{Springer Berlin Heidelberg}, \bibinfo{address}{Berlin, Heidelberg}, \bibinfo{pages}{145--167}.
\newblock
\showISBNx{978-3-642-87825-1}


\bibitem[Veach and Guibas(1995b)]%
        {veach1995combine}
\bibfield{author}{\bibinfo{person}{Eric Veach} {and} \bibinfo{person}{Leonidas~J. Guibas}.} \bibinfo{year}{1995}\natexlab{b}.
\newblock \showarticletitle{Optimally combining sampling techniques for Monte Carlo rendering}. In \bibinfo{booktitle}{\emph{Proceedings of the 22nd Annual Conference on Computer Graphics and Interactive Techniques}} \emph{(\bibinfo{series}{SIGGRAPH '95})}. \bibinfo{publisher}{Association for Computing Machinery}, \bibinfo{address}{New York, NY, USA}, \bibinfo{pages}{419–428}.
\newblock
\showISBNx{0897917014}
\urldef\tempurl%
\url{https://doi.org/10.1145/218380.218498}
\showDOI{\tempurl}


\bibitem[Vorba et~al\mbox{.}(2014)]%
        {vorba2014gmm}
\bibfield{author}{\bibinfo{person}{Ji\v{r}\'{\i} Vorba}, \bibinfo{person}{Ond\v{r}ej Karl\'{\i}k}, \bibinfo{person}{Martin \v{S}ik}, \bibinfo{person}{Tobias Ritschel}, {and} \bibinfo{person}{Jaroslav K\v{r}iv\'{a}nek}.} \bibinfo{year}{2014}\natexlab{}.
\newblock \showarticletitle{On-line learning of parametric mixture models for light transport simulation}.
\newblock \bibinfo{journal}{\emph{ACM Trans. Graph.}} \bibinfo{volume}{33}, \bibinfo{number}{4}, Article \bibinfo{articleno}{101} (\bibinfo{date}{July} \bibinfo{year}{2014}), \bibinfo{numpages}{11}~pages.
\newblock
\showISSN{0730-0301}
\urldef\tempurl%
\url{https://doi.org/10.1145/2601097.2601203}
\showDOI{\tempurl}


\bibitem[Yilmazer et~al\mbox{.}(2024)]%
        {yilmazer2024inverse}
\bibfield{author}{\bibinfo{person}{Ekrem~Fatih Yilmazer}, \bibinfo{person}{Delio Vicini}, {and} \bibinfo{person}{Wenzel Jakob}.} \bibinfo{year}{2024}\natexlab{}.
\newblock \showarticletitle{Solving Inverse PDE Problems using Monte Carlo Estimators}.
\newblock \bibinfo{journal}{\emph{Transactions on Graphics (Proceedings of SIGGRAPH Asia)}}  \bibinfo{volume}{43} (\bibinfo{date}{Dec.} \bibinfo{year}{2024}).
\newblock
\urldef\tempurl%
\url{https://doi.org/10.1145/3687990}
\showDOI{\tempurl}


\bibitem[Yu et~al\mbox{.}(2024)]%
        {yu2024diffwos}
\bibfield{author}{\bibinfo{person}{Z. Yu}, \bibinfo{person}{L. Wu}, \bibinfo{person}{Z. Zhou}, {and} \bibinfo{person}{S. Zhao}.} \bibinfo{year}{2024}\natexlab{}.
\newblock \showarticletitle{A Differential Monte Carlo Solver For the Poisson Equation}. In \bibinfo{booktitle}{\emph{ACM SIGGRAPH 2024 Conference Proceedings}}.
\newblock


\bibitem[Zeltner et~al\mbox{.}(2021)]%
        {zeltner2021differentialmc}
\bibfield{author}{\bibinfo{person}{Tizian Zeltner}, \bibinfo{person}{Sébastien Speierer}, \bibinfo{person}{Iliyan Georgiev}, {and} \bibinfo{person}{Wenzel Jakob}.} \bibinfo{year}{2021}\natexlab{}.
\newblock \showarticletitle{Monte Carlo Estimators for Differential Light Transport}.
\newblock \bibinfo{journal}{\emph{Transactions on Graphics (Proceedings of SIGGRAPH)}} \bibinfo{volume}{40}, \bibinfo{number}{4} (\bibinfo{date}{Aug.} \bibinfo{year}{2021}).
\newblock
\urldef\tempurl%
\url{https://doi.org/10.1145/3450626.3459807}
\showDOI{\tempurl}


\bibitem[Zhu et~al\mbox{.}(2021)]%
        {zhu2021conv}
\bibfield{author}{\bibinfo{person}{Shilin Zhu}, \bibinfo{person}{Zexiang Xu}, \bibinfo{person}{Tiancheng Sun}, \bibinfo{person}{Alexandr Kuznetsov}, \bibinfo{person}{Mark Meyer}, \bibinfo{person}{Henrik~Wann Jensen}, \bibinfo{person}{Hao Su}, {and} \bibinfo{person}{Ravi Ramamoorthi}.} \bibinfo{year}{2021}\natexlab{}.
\newblock \showarticletitle{Hierarchical neural reconstruction for path guiding using hybrid path and photon samples}.
\newblock \bibinfo{journal}{\emph{ACM Trans. Graph.}} \bibinfo{volume}{40}, \bibinfo{number}{4}, Article \bibinfo{articleno}{35} (\bibinfo{date}{July} \bibinfo{year}{2021}), \bibinfo{numpages}{16}~pages.
\newblock
\showISSN{0730-0301}
\urldef\tempurl%
\url{https://doi.org/10.1145/3450626.3459810}
\showDOI{\tempurl}


\end{thebibliography}

\clearpage

\clearpage

\clearpage

\includepdf[pages=-]{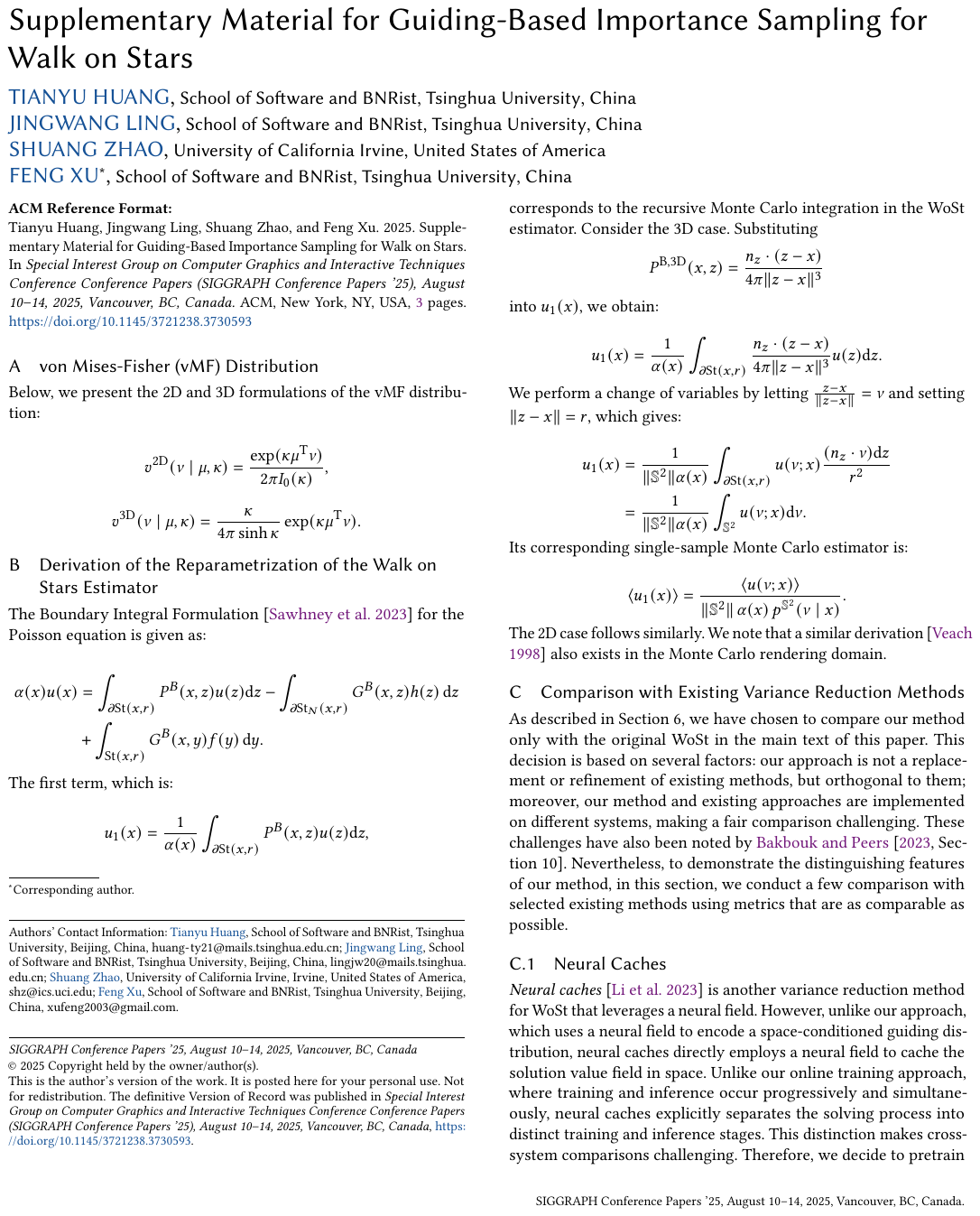}

\end{document}